\documentclass[11pt]{iopart}
\usepackage{graphicx} 
\usepackage{amsmath}
\allowdisplaybreaks
\usepackage{amsgen}
\usepackage{amssymb}
\usepackage{setspace}
\usepackage{setstack}
\usepackage{color}
\usepackage{yfonts}
\usepackage{graphics}
\usepackage{eufrak}
\usepackage[colorlinks=true, allcolors=blue]{hyperref}
\usepackage[colorlinks=true, allcolors=blue]{hyperref}
\begin{document}
\title[]{Compact binary systems in the post-Newtonian limit of gravitational theories with preferred frames}

\author{Oliver Pitt$^{1}$ and Timothy Clifton$^{2}$}
\address{Department of Physics \& Astronomy, Queen Mary University of London, UK.}
\ead{$^1$o.pitt@qmul.ac.uk, $^2$t.clifton@qmul.ac.uk}

\vspace{0.5cm}
\begin{abstract}
Gravitational-wave observations of compact-binary inspirals offer the opportunity to test deviations from general relativity in both strong and weak-field regimes. In previous work, we developed a theory-independent formalism for describing compact-binary dynamics in metric theories of gravity without preferred frames, using the  parametrised post-Newtonian (PPN) couplings together with an additional set of scalar sensitivity parameters that encode violations of the strong equivalence principle. Here, we extend this framework to include preferred-frame effects, of the type that can arise in (for example) vector-tensor theories of gravity. This is achieved by introducing a time-like vector field that identifies the preferred frame, and then by allowing compact body masses to depend on the invariant formed from this field and the body's four-velocity. We calculate the modified equation of motion up to first post-Newtonian order (1PN), where the preferred-frame PPN parameters $\alpha_1$ and $\alpha_2$ acquire contributions from our new vector sensitivities. We further combine scalar and vector sensitivities in a unified formalism, which we validate with an example scalar-vector-tensor theory. 
This approach provides the orbital dynamics required to extend gravitational-wave constraints on PPN and sensitivity parameters to theories of gravity that admit preferred frames, as well as those that do not. 
\end{abstract}

\section{Introduction}

General relativity (GR) has succeeded in passing a broad range of observational tests, including measurements of binary pulsar timings, solar system experiments, and gravitational wave observations \cite{freire2024gravity,will2018theory,abac2026gwtc}. At the same time, the efforts to understand the late-time acceleration of the cosmos and the absence of a valid quantum theory of gravity motivate a vigorous search for alternatives \cite{clifton2012modified}. A powerful approach to this search is to calculate physical observables using theory-independent methods, with one of the most successful frameworks being the parametrised post-Newtonian (PPN) formalism \cite{will2018theory}.

The PPN formalism was originally designed for Solar System tests of metric theories of gravity, and has become a standard tool for constraining them using observations from the weak-field and slow-motion regime \cite{will2018theory}.  To make use of the formalism's constraining power, much effort has been spent extending it beyond the isolated astrophysical systems for which it was originally built \cite{anton2022momentum,thomas2023scale,thomas2024constraining}. Developing the PPN formalism so that it can be applied in compact binary systems was the  aim of our previous work, in Reference  \cite{pitt2025constraining}, where we demonstrated how the phase of a gravitational wave signal emitted from a compact binary system depends on both the PPN parameters and an additional set of theory-independent sensitivity parameters, introduced to accommodate the strong field effects, as expected in compact binaries in such theories. 

A further motivation for a theory-independent description of gravitational waves from compact-binary inspirals is that most existing tests of general relativity in this field are predominantly phenomenological \cite{abac2026gwtc}. That is, one typically constrains parametrised deformations of the waveform phase rather than the coupling parameters responsible for generating the gravitational field itself. Such tests are deliberately theory agnostic, but their interpretation in terms of a specific modified theory requires an additional mapping between the phenomenological deformation parameters and the theory's couplings. Our aim is not to remove this modelling step entirely, but to place the parametrization closer to the weak-field quantities used in existing gravitational tests.

A theory-independent approach rooted in the PPN formalism starts from quantities with clear physical meaning in the weak-field regime. For example, the parameter $\gamma$ measures the amount of spatial curvature produced per unit mass, while $\beta$ parametrizes non-linear gravitational self-interaction, and $\alpha_1$ and $\alpha_2$ encode preferred-frame effects.  The PPN framework provides a common language for a broad class of metric theories, allowing deviations from GR to be organised in terms of parameters that encode identifiable physical effects. Extending such a framework to compact-binary inspirals therefore offers the prospect of constraints that are more transparent, more readily comparable with existing weak-field bounds, and more directly connected to underlying gravitational physics than current phenomenological inspiral tests alone.

One of the restrictions of our original work was that it was limited to theories of gravity that do not admit preferred frames of reference. This was sufficient to match the equation of motion and radiative degrees of freedom for e.g. scalar-tensor theories, but is not sufficient for e.g. vector-tensor theories. Addressing this deficit is the aim of the present work, which is motivated by some approaches to quantum gravity generating preferred-frame effects in the gravitational sector \cite{moffat2010lorentz}. Examples include vector-tensor theories theories of gravity, such as Einstein-\AE ther theory, which provide covariant low-energy effective descriptions of some such theories of quantum gravity \cite{eling2006einstein}.

In Section \ref{sc:background}, we lay out the necessary theoretical ingredients for our approach, namely the PPN formalism and compact body sensitivities. Section \ref{sc:VT} then introduces a post-Newtonian vector field and uses it to define a preferred frame in which to construct a theory-independent equation of motion for compact bodies with sensitivities. In Section \ref{sc:SVT} we show how this approach can be combined with our previous work in Reference \cite{pitt2025constraining}, which we apply to an example scalar-vector-tensor theory in Section \ref{sc:FullTheory}. We discuss our results in Section \ref{sc:Diss}, and finish with some concluding remarks in Section \ref{sc:Conc}. Auxiliary material may be found in the appendices, and we work with units in which $G=c=1$ at the present cosmological time.


\section{Theoretical Background}
\label{sc:background}

In this section, we will discuss the two main theoretical ingredients that form the backbone of our formalism: the PPN formalism and sensitivities. 

\subsection{The parametrised Post-Newtonian Formalism}

The PPN formalism was originally built to describe metric theories of gravity in a slow-motion and weak-field expansion on a Minkowski background \cite{will2018theory}. The smallness parameter of the expansion is $\eta \sim v$, where $v \ll c$ is understood as the 3-velocity of matter in the system. Additionally, it is assumed that all gravitational fields vary on time scales in accordance with this velocity, such that the time derivatives of such fields pick up an order of smallness compared with spatial derivatives, i.e.  ${\partial_t} \sim \eta \, {\partial_x}$.
   
Through the Euler equation, the Newtonian potential is then of size
\begin{equation}
    U \sim v^2 \sim \eta^2 \, ,
    \label{eq:EulPN}
\end{equation}
where $\nabla^2 U = - 4 \pi \, \rho^*$, where $\rho^*=\rho + \frac{1}{2} v^2 +3 \gamma U$ is the `conserved density', and where $\gamma$ is the PPN parameter introduced below (not the Lorentz factor). In the post-Newtonian dynamics for massive bodies the leading order contributions (0PN) are proportional to $\mathcal{O}(\eta^2)$, the next-to-leading order terms (1PN) go as $\mathcal{O}(\eta^4)$ and so on for higher orders of even powers of $\eta$. We take this approximation to be valid for the early inspiral of compact binary mergers. In the PPN gauge, defined by $g_{ij} \propto \delta_{ij}$ and  $g_{00}$ having no dependence on the $\Phi_5$ potential \cite{will2018theory},  for point masses and in semi-conservative theories of gravity, we have the following metric components \cite{will1972conservation}:
\begin{align}
    g_{00} &= -1 +2 \alpha U - 2 \beta U^2 +(2 \alpha^2 -4 \beta ) \sum_{A,B} \frac{m_{{\rm }A} m_{{\rm } B}}{r_A r_{AB}} + (2 \gamma + \alpha) \sum_A \frac{m_{{\rm } A} v_A^2}{r_A} +\mathcal{O}(\eta^5) \label{ppn1} \\
    g_{0i} &= - \frac{1}{2} (4 \gamma +4 \alpha + \alpha_1 ) \sum_A \frac{m_{{\rm }A} v_A^i}{r_A} -\frac{1}{2} (\alpha +\alpha_2 ) \sum_A \frac{m_{{\rm }A}}{r_A} \, (\vec{v}_A \cdot \vec{n}_A) \, r_A^i +\mathcal{O}(\eta^4) \label{ppn2} \\[4pt]
    g_{ij} &= (1+2 \gamma U) \delta_{ij} +\mathcal{O}(\eta^3)\, , \label{ppn3}
\end{align}
where $U=\sum_A m_{{\rm }A}/r_A$ is the Newtonian gravitational potential, $\vec{v}_A$ is the 3-velocity of body $A$ and $\vec{r}_{AB} \equiv \vec{r}_{A}-\vec{r}_B$. The parameter $\alpha$ has been included to accommodate the possible cosmological time-variation of Newton's constant \cite{sanghai2017parameterized} and we have omitted the Whitehead term. We note the inclusion of $\alpha$ propagates through to 1PN order, so we must also include it at that order to maintain consistent definitions of the other PPN parameters. 

We take this metric to be sufficient to calculate the post-Newtonian geodesic motion of point particles for theories of gravity that would be semi-conservative in the absence of sensitivities \cite{clifton2026post}. In the following sections we will generalise this description so that we can use it for compact bodies.

\subsection{Sensitivities}

Modified theories of gravity often possess additional gravitational fields that couple to matter and the metric. This coupling leads to violations of the Strong Equivalence Principle (SEP), which means that a body's internal structure can affect its motion. In theories without the SEP, and in particular in compact objects like neutron stars or black holes, we do not necessarily expect bodies to follow geodesics of space-time. In 1975, Eardley developed an effective treatment for this behaviour by modelling bodies as point particles with environmentally-dependent masses \cite{eardley1975observable}. Originally developed for scalar-tensor theories of gravity, Eardley distinguished between the inertial mass of a body, $m_0$, and the gravitational mass dependent on the scalar field, $m=m(\phi)$. One can generalise Eardley's procedure by making the mass a function of arbitrary external fields $\psi_a$ \cite{taherasghari2022modified}, which results in the following matter action:
\begin{equation}
\label{eq:eardley equation}
    I_m = - \sum_A \int m_A \left( \psi_a [{\bf x}_A(\tau_A)] \right) d\tau_A \, ,
\end{equation}
where e.g. ${\bf x}_A(\tau_A)$ parametrizes the world-line of the $A$th body in terms of its proper time, $\tau_A$. Combining this action with the appropriate gravitational action leads to field equations with a mass sensitive to each of these additional fields. This sensitivity is parametrised through derivatives of the mass, as follows:
\begin{equation}
s_A^{a} \equiv \frac{\partial \ln m_A}{\partial \psi_a} \Bigg\vert_0\,
\qquad {\rm and } \qquad
s_A^{\prime \,ab} \equiv \frac{\partial^2 \ln m_A}{\partial \psi_a \, \partial \psi_b} \Bigg\vert_0\, ,
\end{equation}
where  $m_{A}\vert_0=m_A\big(\psi_a^{(0)}\big)$. These sensitivity parameters are in place to encode the way in which the additional degrees of freedom in modified theories of gravity affect the internal structure of a body and therefore its motion. They mean that the mass of a body can be written as 
\begin{align}
    m_A(\psi_a) 
    &= m_{A}\vert_0 \left(1+ \sum_a s_A^a \, \delta \psi_a +\frac{1}{2} \sum_{a,b} s_A^{\prime \,ab} \, \delta \psi_a \, \delta \psi_b  \right) + \mathcal{O}(\delta \psi_a^3) \, .
\end{align}
In the following section we will see how this idea can be extended to generate theory-independent sensitivities in cases where there exists a preferred frame of reference.

\section{Theory-Independent Equations of Motion with Vector Sensitivities}
\label{sc:VT}

Here we set out how to construct theory-independent equations of motion to describe the motion of compact bodies to 1PN in the presence of sensitivity to vector fields that pick out a preferred time-like direction at each location in space-time; a situation we will describe as the body having a `vector sensitivity'.

\subsection{A Theory-Independent Vector Field}

In our previous work,  we allowed the mass of compact bodies to depend upon the scalar Newtonian and post-Newtonian gravitational potentials \cite{pitt2025constraining}. To extend our theory-independent approach to cases with preferred-frame effects, we define a unit time-like post-Newtonian vector field, $K^\mu=(K^0,K^i)$, to which our masses should be sensitive:
\begin{align} \label{kvec}
K^{0} 
&= g_{00}^{-1/2}, 
\qquad  \qquad
K^{i} 
= C^{+} V^i
 + C^{-} W^i
\end{align}
with 
\begin{align}
    V^i=\sum_A \frac{ m_A}{r_A} v^i_A   \qquad   \text{and}\qquad  W^i=\sum_A  \frac{ m_A}{r_A} 
   \big( \vec{v}_A \cdot \vec{n}_A \big) n_A^{i}\,.
\end{align}
Here $C^{-}$ and $C^{+}$ are theory-independent degrees of freedom, $v^i_A$ is particle A's 3-velocity, $r_A$ is the distance to body A and $\vec{n}_A=\vec{r}_A/|r_A|$ is the associated unit vector. Note that because $K^\mu$ is taken to be a normalised time-like vector we have that $K^\mu K_\mu =-1$, which fixes the definition of $K^0$ to the required order. The existence of such a vector field provides a way to specify a preferred frame within a covariant formalism, without specifying anything else about the theory of gravity, as required for our purposes. 

An important consideration is that the spatial part of the vector, $K^i$, depends only on potentials of order $v^3/c^3$. In principle, the vector field might possess a lower-order component or velocity. This has not been neglected; rather, we assume we are working in coordinates picked out by the field; i.e. such that $w^i=0$, where $w^i$ is understood to be the leading-order component of the 3-velocity of the preferred frame. As this is the only lower-order quantity available that has the required index structure, we are led to the form in Equation (\ref{kvec}). We can re-introduce the bodies' dependence on $w^i$ by boosting our final equation of motion, if this is required (see the appendix for details). Under the assumption that the spin of a body can be neglected, the only available invariant upon which the body's internal structure might depend is then $\gamma=-K^\mu u_\mu$, where $u_\mu$ is the body's four-velocity. This is the Lorentz factor for the boost between the body's rest frame and the frame picked out by $K^\mu$, which leads to a matter action of the form \cite{foster2007strong,taherasghari2023compact} 
\begin{align}
\label{eq:VECaction}
S_M 
&= - \sum_A \int 
    m_A(\gamma) \, 
    d\tau_A,
\end{align}
 where $\tau_A$ is the proper time along body A's world-line, and $m_A$ is the mass of the test particle we are using to model it. From Equation (\ref{eq:VECaction}) we obtain the following equations of motion: 
 \begin{align}
\label{eq:geo1}
u_A^{\nu} \nabla_{\nu} \big[ 
    m_A u_{A\alpha} 
    + m'_A K^{\mu} \big( g_{\mu\alpha} + u_{A\mu} u_{A\alpha} \big) 
\big]
&= m'_A u_{A\mu} \nabla_{\alpha} K^{\mu}\, ,
\end{align}
where $m'_A={\partial m_A}/{\partial \gamma}$ and where $u_A^{\mu}$ is the four-velocity of test body A (see the appendix for a full derivation).

\subsection{Equations of Motion with Vector Sensitivities}

Let us now define a post-Newtonian series expansion for a sensitive mass: 
\begin{align}
m_A(\gamma) 
&= m_A \left[ 
    1 + s_A (\gamma - 1) 
    + \tfrac{1}{2} a_{sA} (\gamma - 1)^2 
+ \mathcal{O}((\gamma-1)^3) \right] \,,
\end{align}
where
\begin{align}
s_A &\equiv 
\left( 
    \frac{d \ln m_A(\gamma)}{d \ln \gamma}
\right)\Bigg\vert_{\gamma = 1} \qquad {\rm and} \qquad  s'_A \equiv 
\left( \frac{d^{2} \ln m_A(\gamma)}{d (\ln \gamma)^{2}} 
\right)\Bigg\vert_{\gamma = 1}
\end{align}
and where $a_{sA} \equiv s_A^{2} - s_A + {s'_A}$. We are now in a position to write Equation (\ref{eq:geo1}) to Newtonian order as
\begin{align}
    \boldsymbol{{a}_1}= -\frac{\alpha}{1-s_1}\frac{m_2}{r_{12}^2}\boldsymbol{n},
\end{align}
where the replacements $1\rightarrow 2$ and $\boldsymbol{n}\rightarrow \boldsymbol{-n}$ give the equation of motion for the other body in the system. We note that this method has produced the complete Newtonian equation for both bodies, which is in contrast to theories with scalar sensitivities only. In that case, the absent contribution of body 1's sensitivity had to be deduced in the equation of motion for body 2 by requiring that the passive and active gravitational masses of body 1 are equal. Here, it would appear to be better to state that it is not the gravitational masses that are sensitive, but rather the inertial mass. This is more clearly seen if the equation is written as 
\begin{align}
    m_1\big(1-s_1\big)\boldsymbol{a}_1 = -\alpha\frac{m_1m_2}{r^2}\boldsymbol{n}.
\end{align}
For clarity, we can follow Reference \cite{foster2007strong} and re-scale the mass of body 2 in the equation of motion for body 1 so that the sensitive prefactor is symmetric, using $m_2\rightarrow \tilde m_2/(1-s_2)$, such that 
\begin{equation}
    \boldsymbol{a_1}= -\frac{\alpha}{(1-s_1)(1-s_2)}\frac{\tilde m_2}{r^2}\boldsymbol{n} \, .
\end{equation}
We can now proceed to determine Equation (\ref{eq:geo1}) to post-Newtonian order for a sensitive body orbiting a non-sensitive body, which gives
\begin{equation} \label{eq:ppN+Sen1}
\begin{split}
  \boldsymbol{a}_1 = &-\frac{\alpha}{1-s_1}\frac{m_2}{r^2}\boldsymbol{n}+\frac{ m_2}{(1-s_1)r^2}\boldsymbol{n}\left[ \left( 2\alpha\gamma+2\beta\right) \frac{m_2}{r} \right.\\ 
    & \left. + \frac{1}{2}\left(\alpha\left(4\gamma +4\alpha+\alpha_1 -s_1C^+\right)+4\beta - 2\alpha^2\right)\frac{m_1}{r}  \right. \\
    &\left. +\frac{1}{2} \left(4\gamma+4\alpha+\alpha_1-s_1C^+ \right)\left(\boldsymbol{v}_1 \cdot \boldsymbol{v}_2 \right) -\frac{1}{2}\left( 2\gamma+2\alpha+\alpha_2-2s_1C^- \right)v_2^2\right. \\ 
   &\left. -\left(\frac{2\gamma-s_1(2\gamma+\alpha)}{2}-\frac{\alpha (a_{s1})}{2(1-s_1)} \right) v_1^2
     +\frac{3}{2}
     \left(\alpha+\alpha_2-2s_1C^-\right)(\boldsymbol{n}\cdot\boldsymbol{v}_2)^2  \right] \\
     &+\frac{ m_2}{(1-s_1)r^2}\boldsymbol{n} \cdot \left[ \left( 2\gamma(1-s_1)+\alpha(2-s_1) +\frac{\alpha(a_{s1})}{1-s_1}\right) \boldsymbol{v_1}-\left( 2\gamma+\alpha \right)(1-s_1)\boldsymbol{v}_2 \right] \boldsymbol{v}_1\\
    &-\frac{1}{2}\frac{ m_2}{(1-s_1)r^2}\boldsymbol{n}\cdot \left[ \left( 4\gamma+4\alpha+\alpha_1-s_1C^+ \right) \boldsymbol{v}_1- \left( 4\gamma+2\alpha+\alpha_1-2\alpha_2+s_1(4C^--C^+)  \right) \boldsymbol{v}_2 \right] \boldsymbol{v}_2 \, .
\end{split}
\end{equation}
As was the case for theories that admitted sensitivity to scalar potentials, we can generalise this situation by using the symmetry properties of the modified Einstein-Infeld-Hoffmann (EIH) parameters, which allow us to determine the equations of motion in the case where both bodies are sensitive \cite{will2018testing}. These are
\begin{align}
    \mathcal{G}_{ab} = \mathcal{G}_{(ab)}, \quad \mathcal{C}_{ab} = \mathcal{C}_{(ab)}, \quad \mathcal{E}_{ab} = \mathcal{E}_{(ab)}, \quad {\rm and } \quad \mathcal{D}_{abc} = \mathcal{D}_{a(bc)} \, ,
\end{align}
with $\mathcal{B}_{ab}$ having no particular symmetry.
To have consistently defined modified EIH parameters, we work with the rescaled masses of the bodies, i.e. $m_a\rightarrow \tilde m_a/(1-s_a)$. The corresponding EIH parameters are set out in Table \ref{tab:VTSen} (see the appendix for further detail on the EIH Lagrangian and its associated parameters). 

The parameters $\mathcal{C}_{ab}$ and $\mathcal{E}_{ab}$ are in contrast to the other EIH parameters in Table \ref{tab:VTSen} as they contain sensitivities that appear in the gravitational masses. Gravitational sensitivities must be `dualled' when generalising them to describe the motion of two sensitive bodies, which can be achieved by demanding that the active and passive gravitational masses of a sensitive body be equivalent. Requiring that the dualled version be linear in the individual sensitivities, and symmetric under interchange, gives the general form 
\begin{align}
    s_{\{12\}}= s_{(1)}+s_{(2)}+\frac{1}{C}s_{(1)}s_{(2)},
\end{align}
where $C$ is an arbitrary constant, which should be expressible as a function of the parameters of whatever theory is being considered. In Table \ref{tab:VTSen} the parameters $D^+$ and $D^-$ fulfil this role for $\mathcal{C}_{ab}$ and $\mathcal{E}_{ab}$, respectively. 
\begin{table}[t]
    \centering
    \begin{tabular}{|c|c|}
        \hline
        EIH Parameter & Form for Two Sensitive Bodies $\phantom{\Big(}$ \\[5pt] \hline
        $\mathcal{S}_{ab}$ & $\frac{1}{(1-s_a)(1-s_b)} \phantom{\Big(}$ \\[5pt] \hline
        $\mathcal{G}_{ab}$ & $\alpha\mathcal{S}_{ab}  \phantom{\Big(}$ \\[5pt] \hline
        $\mathcal{A}_a$&$(a_{sa})/(1-s_a)$ \\[5pt] \hline
        $\mathcal{B}_{ab}$ & $\frac{1}{3}\mathcal{S}_{ab}\left(2\gamma+\alpha\right)(1-s_a) \phantom{\Big(}$  \\ \hline
        $\mathcal{D}_{abc}$ & $\mathcal{S}_{ab}\mathcal{S}_{ac}(2\beta-\alpha^2)(1-s_a)\phantom{\Big(}$   \\[5pt] \hline
        $\mathcal{E}_{ab}$&$\mathcal{S}_{ab}(\alpha_2-(s_a+s_b)(2C^-)+s_as_bD^-)$\\[5pt] \hline
        $\mathcal{C}_{ab}$& $\mathcal{S}_{ab}(\alpha_1-\alpha_2 -(C^+-2C^-+2\gamma+\alpha)(s_a+s_b)+(D^+-2D^-)s_as_b)$\\[5pt] \hline
    \end{tabular}
    \caption{The  modified EIH parameters for theories with preferred frame effects, when both bodies are sensitive. Note that the parameters $D^-$ and $D^+$ are arbitrary constants associated with the sensitivities of the two bodies}
    \label{tab:VTSen}
\end{table}
We can now write the theory-independent equations of motion for two sensitive bodies as follows, where the gravitational sensitivities present in $\mathcal{C}_{ab}$ and $\mathcal{E}_{ab}$ combine neatly with the PPN parameters $\alpha_1$ and $\alpha_2$, as in  Table \ref{tab:EIH2}:
\begin{equation}\label{eq:ppN+Sen2}
\begin{split}
  \boldsymbol{a}_1 = &-\alpha\frac{ m_2}{(1-s_2)(1-s_1)r^2}\boldsymbol{n}
  +\frac{ m_2}{(1-s_1)(1-s_2)r^2}\boldsymbol{n}\left[ \left( 2\alpha\gamma+2\beta\right) \frac{m_2}{(1-s_2)r} \right.\\ 
    & \left. + \frac{1}{2}\left(\alpha\left(4\gamma +4\alpha+\tilde\alpha_1 \right)+4\beta - 2\alpha^2\right)\frac{m_1}{(1-s_1)r}  \right. \\
    &\left. +\frac{1}{2} \left(4\gamma+4\alpha+\tilde\alpha_1 \right)\left(\boldsymbol{v}_1 \cdot \boldsymbol{v}_2 \right) -\frac{1}{2}\left( 2\gamma+2\alpha+\tilde\alpha_2 \right)v_2^2\right. \\ 
   &\left. -\left(\frac{2\gamma-s_1(2\gamma+\alpha)}{2}-\frac{\alpha (a_{s1})}{2(1-s_1)} \right) v_1^2
     +\frac{3}{2}
     \left(\alpha+\tilde\alpha_2\right)(\boldsymbol{n}\cdot\boldsymbol{v}_2)^2  \right] \\
     &+\frac{ m_2}{(1-s_1)(1-s_2)r^2}\boldsymbol{n} \cdot \left[ \left( 2\gamma(1-s_1)+\alpha(2-s_1) +\frac{\alpha(a_{s1})}{1-s_1}\right) \boldsymbol{v}_1-\left( 2\gamma+\alpha \right)(1-s_1)\boldsymbol{v}_2 \right] \boldsymbol{v}_1\\
    &-\frac{1}{2}\frac{ m_2}{(1-s_1)(1-s_2)r^2}\boldsymbol{n}\cdot \left[ \left( 4\gamma+4\alpha+\tilde\alpha_1\right) \boldsymbol{v}_1- \left( 4\gamma+2\alpha+\tilde\alpha_1-2\tilde\alpha_2  \right) \boldsymbol{v}_2 \right] \boldsymbol{v}_2 \, , 
\end{split}
\end{equation}
and with $\boldsymbol{a_2}= \{1\rightleftharpoons2;\boldsymbol{n}\rightarrow\boldsymbol{-n}\}$ giving the equation of motion for body 2.

\begin{table}[t!]
    \centering
    \begin{tabular}{|c|c|}
        \hline
        Two-body compact&explicit\\
        \hline
       $\tilde\alpha_1$&$\alpha_1-(C^++2\gamma+\alpha)(s_1+s_2)+D^+s_1s_2$\\
       \hline
    $\tilde\alpha_2$&$\alpha_2-2C^-(s_1+s_2)+2D^-s_1s_2$\\
    \hline
    \end{tabular}
    \caption{Two body $\alpha_1$ and $\alpha_2$ parameters including the effects of vector sensitivities.}
    \label{tab:EIH2}
\end{table}
Equation (\ref{eq:ppN+Sen2}) is sufficient to accommodate the motion of compact objects in vector-sensitive theories; in particular, it aligns with the 1PN equations obtained in Refs. \cite{foster2007strong,taherasghari2023compact} for the particular case of Einstein-\AE ther theories. We will demonstrate this explicitly when we consider an example scalar-vector-tensor theory, below.

\section{Theory-Independent Equations of Motion with both Scalar and Vector Sensitivities}
\label{sc:SVT}

We now seek to find equations of motion that can simultaneously accommodate theories that exhibit both scalar and vector sensitivities.

\subsection{Including Scalar Sensitivities}

We will now allow the masses of our bodies to depend additionally on the following scalar potentials: 
\begin{align}
\label{psiA}
\{ U, \Phi_1, \Phi_2 , \ddot X \}
=
\left\{ \sum_A \frac{m_{{\rm G}A}}{r_A} ,\sum_A \frac{ m_{{\rm G}A} v_A^2}{r_A},
\sum_{A, B\neq A} \frac{ m_{{\rm G}A}m_{{\rm G}B}}{r_A r_{AB}},
\sum_{A}\frac{ m_{{\rm G}A} v_A^2}{r_A}- \frac{ m_{{\rm G}A}}{r_A^3}
(\textbf{v}_A\cdot \textbf{r}_{A})^2 
\right\}.
\end{align}
This results in the equations of motion
\begin{align}
\label{eq:geo2}
u_A^{\nu} \nabla_{\nu} \big[ 
    m_A u_{A\alpha} 
    + \frac{\partial m_A}{\partial \gamma} K^{\mu} \big( g_{\mu\alpha} + u_{A\mu} u_{A\alpha} \big) 
\big]
&= \frac{\partial m_A}{\partial \gamma} u_{A\mu} \nabla_{\alpha} K^{\mu}+\sum_i\frac{\partial m_A}{\partial \psi_i}\nabla_{\alpha}\psi,
\end{align}
where $\psi_i$ represents the additional scalar potentials, and the masses now have the dependence $m_A=m_A(\gamma,\psi_i)$, which possesses the following series expansion:
\begin{equation}
\label{eq:masstest}
    m(\psi_i,\gamma)= m_0+\frac{\partial m}{\partial\psi_i}\psi_i+\frac{\partial m}{\partial \gamma}(\gamma-1) + \frac{1}{2}\frac{\partial^2m}{\partial^2\psi_i}\psi_i^2 + \frac{1}{2}\frac{\partial^2m}{\partial^2\gamma}(\gamma-1)^2 + \frac{\partial^2m}{\partial\psi_i\partial\gamma}\psi_i(\gamma-1)+ \ldots \, ,
\end{equation} 
where 
\begin{equation}
\begin{aligned}
    &s_{\Phi_1} \equiv \frac{d \ln m}{d\Phi_1}, \quad s_{\Phi_2} \equiv \frac{d \ln m}{d\Phi_2}, \quad s_{\ddot X} \equiv \frac{d \ln m}{d\ddot X} , \quad s'_{U} \equiv \frac{d^2 \ln m}{dU^2} \, \\
    &s_{\gamma U}\equiv \frac{\partial^2\ln m}{\partial \gamma \partial U}, \quad    {\rm and}\quad s^{(1)}_{\gamma U}\equiv \left(1-s_\gamma^{(1)}\right)\Sigma^{(1)}_{\gamma U}\, .
\end{aligned}
\end{equation}
\begin{table}[h]
\centering
\resizebox{\textwidth}{!}{%
\begin{tabular}{|c|c|}
\hline
\textbf{EIH parameter} & \textbf{Body 1 sensitive} \\
\hline
$\mathcal{G}_{12}$ & $\displaystyle\frac{\alpha - s^{(1)}_U}{1 - s^{(1)}_\gamma}$ \\[10pt]
\hline
$\mathcal{A}_1$ & $\displaystyle\frac{a^{(1)}_{s\gamma}}{1 - s^{(1)}_\gamma}$ \\[10pt]
\hline
$\mathcal{B}_{12}$ & $\displaystyle\frac{1}{3}\left(2\gamma + \alpha + s^{(1)}_U - \Sigma_{\gamma U}^{(1)}\right)$ \\[10pt]
\hline
$\mathcal{B}_{21}$ & $\displaystyle\frac{1}{3}\left(2\gamma + \alpha - 2s^{(1)}_{\Phi_1}\right)\frac{1}{1 - s^{(1)}_\gamma}$ \\[10pt]
\hline
$\mathcal{D}_{122}$ & $\displaystyle\left(2(\beta - \alpha^2) + \left(\alpha - s^{(1)}_U\right)^2 + s'^{(1)}_U\right)\frac{1}{1 - s^{(1)}_\gamma}$ \\[10pt]
\hline
$\mathcal{D}_{211}$ & $\displaystyle\left(2\beta - \alpha^2 - f(s^{(1)}_U) + s^{(1)}_{\Phi_2}\right)\frac{1}{1 - s^{(1)}_\gamma}$ \\[10pt]
\hline
$\mathcal{C}_{12}$ & $\displaystyle\left(\alpha_1 - \alpha_2 - \left(C^+ - 2C^- + 2\gamma + \alpha + s^{(1)}_U - \Sigma^{(1)}_{\gamma U}\right)s^1_\gamma + 2s^{(1)}_{\Phi_1} + 2s^{(1)}_{\ddot{X}}\right)\frac{1}{1 - s^{(1)}_\gamma}$ \\[10pt]
\hline
$\mathcal{E}_{12}$ & $\displaystyle\left(\alpha_2 - 2C^- s^1_\gamma - 2s^{(1)}_{\ddot{X}} + s^{(1)}_U\right)\frac{1}{1 - s^{(1)}_\gamma}$ \\[10pt]
\hline
\end{tabular}
}
\caption{The corresponding EIH parameters for our equation of motion with both scalar and vector sensitivities. }
\label{tab:EIH}
\end{table}
Proceeding as before, we calculate Equation (\ref{eq:geo2}) to first post-Newtonian order to obtain an equation of motion with both vector and scalar sensitivities:
\begin{equation} \label{eq:ppN+combdsen}
\begin{split}
  \boldsymbol{a}_1 = &-\frac{\alpha-s^{(1)}_U}{1-s^{(1)}_{\gamma}}\frac{m_2}{r^2}\boldsymbol{n}+\frac{ m_2}{(1-s^{(1)}_{\gamma})r^2}\boldsymbol{n}\left[ \left( 2\alpha\gamma+2\beta-2s^{(1)}_U(\alpha+\gamma)-\Sigma^{(1)}_{\gamma U}(\alpha-s^{(1)}_U)+s^{(1)\prime}_U\right) \frac{m_2}{r} \right.\\ 
    & \left. + \frac{1}{2}\left((\alpha-s^{(1)}_U)\left(4\gamma +4\alpha+\alpha_1 -s^{(1)}_{\gamma}C^+\right)+4\beta - 2\alpha^2- 2 f \big( s_U^{(1)} \big)  +2s^{(1)}_{\Phi_2}\right)\frac{m_1}{r}  \right. \\
    &\left. +\frac{1}{2} \left(4\gamma+4\alpha+\alpha_1-s^{(1)}_{\gamma}C^+ \right)\left(\boldsymbol{v}_1 \cdot \boldsymbol{v}_2 \right) -\frac{1}{2}\left( 2\gamma+2\alpha+\alpha_2-2s^{(1)}_{\gamma}C^- -2s_{\ddot X}^{(1)}-2s^{(1)}_{\Phi_1}\right)v_2^2\right. \\ 
   &\left. -\frac{1}{2}\left((1-s_\gamma^{(1)})(2\gamma+\alpha+s^{(1)}_U-\Sigma_{\gamma U}^{(1)} )-(\alpha -s_U^{(1)})\left(1+\frac{a_{s1}}{1-s^{(1)}_{\gamma}}\right) \right) v_1^2 \right.\\
    &\left. +\frac{3}{2}
     \left(\alpha+\alpha_2-2s^{(1)}_{\gamma}C^--2s_{\ddot X}^{(1)}\right)(\boldsymbol{n}\cdot\boldsymbol{v}_2)^2  \right] \\
     &+\frac{ m_2}{(1-s^{(1)}_{\gamma})r^2}\boldsymbol{n} \cdot \left[ \left((1-s_\gamma^{(1)})(2\gamma+\alpha+s^{(1)}_U-\Sigma_{\gamma U}^{(1)} )+(\alpha -s_U^{(1)})\left(1+\frac{a_{s1}}{1-s^{(1)}_{\gamma}}\right) \right) \boldsymbol{v_1} \right.\\
     &\left.-\left( 2\gamma+\alpha+s^{(1)}_U-\Sigma_{\gamma U}^{(1)} \right)(1-s^{(1)}_{\gamma})\boldsymbol{v}_2 \right] \boldsymbol{v}_1\\
    &-\frac{1}{2}\frac{ m_2}{(1-s^{(1)}_{\gamma})r^2}\boldsymbol{n}\cdot \left[ \left( 4\gamma+4\alpha+\alpha_1-s^{(1)}_{\gamma}C^+ \right) \boldsymbol{v}_1- \right.\\
    &\left. \left( 4\gamma+2\alpha+\alpha_1-2\alpha_2+s^{(1)}_{\gamma}(4C^--C^+) +4s_{\ddot X}^{(1)} \right) \boldsymbol{v}_2 \right]\boldsymbol{v}_2 \, .
\end{split}
\end{equation}
The reader may note that here we have chosen to use a different set of scalar sensitivities than the ones we used previously in Reference \cite{pitt2025constraining}, which were \{$U,\Phi_1,\Phi_2,\Phi_6$\} (i.e. instead of $\Phi_6$ we use $\ddot X$). This is a slightly more convenient choice of variables as it tends to match scalar-tensor theories more closely, leading to $s_{\Phi_1}$ cancelling with $s_{\ddot X}$ rather than just vanishing on its own.



\subsection{Adapting to Two Sensitive Bodies}

In a theory-independent approach, the dependence of the metric on sensitivities is not easily specified. However, at the level of the equations of motion one can use the symmetry properties of the EIH parameters to deduce the contribution to the equations of motion from the sensitivity parameters in the metric. 



We note that in nearly all cases the requirement of having consistent definitions for the modified EIH parameters (and therefore our parameters), in the presence of a second sensitive body, has quite different consequences for the sensitivity parameters that appear in the gravitational and inertial masses. That is, the sensitivities in the gravitational masses must all be dualled to include the effects of the second body, while the sensitivities in the inertial masses do not require any such process. An exception to this  is given by $\Sigma_{\gamma U}^{(1)}$, which is a combined scalar/vector sensitivity and therefore is a combination of both gravitational and inertial mass sensitivities.  

The $\Sigma_{\gamma U}^{(1)}$ parameter features in the definition of $\mathcal{B}_{12}$, whose properties we can use to see how it should behave. In the absence of vector sensitivities, $\mathcal{B}_{12}$ is symmetric under the exchange of indices, and takes the form 
\begin{equation}
\mathcal{B}_{(12)}= \frac{1}{3} \left( 2 \gamma + \alpha + s_{{\rm N}}^{\{12\}} \right) \, .
\end{equation}
Implicit in this definition is the condition $s_{\Phi_1}=-\frac{1}{2}s_U$, which is necessary for $\mathcal{B}_{(12)}$ to maintain a consistent definition when identified through the equations of motion. If one assumes that $s_{\Phi_1}$ retains this property in the combined approach, when $\mathcal{B}_{12} \neq \mathcal{B}_{21}$, then one arrives at 
\begin{equation}
\mathcal{B}_{ab} = \frac{1}{3\left(1-s_\gamma^{(b)}\right)}\left(2\gamma+\alpha+\tilde\Sigma_{\gamma U}^{\{ab\}}\right)\, ,
\end{equation}
where
\begin{align}
    \tilde\Sigma^{(a)}_{\gamma U}\equiv& s^{(a)}_U-\Sigma_{\gamma U}^{(a)}\qquad {\rm and} \qquad
    \tilde\Sigma^{\{ab\}}_{\gamma U}\equiv\tilde\Sigma^{(a)}_{\gamma U} +s_U^{(b)}+\frac{1}{c_{N}}\tilde\Sigma^{(a)}_{\gamma U}s_U^{(b)}.
\end{align}
This conclusion can also be reached if one considers a theory of gravity in which matter is directly coupled to a vector and scalar field, such that
\begin{equation}
\label{eq:masstest}
    m(\phi,\gamma)= m_0+\frac{\partial m}{\partial\phi}\phi+\frac{\partial m}{\partial \gamma}(\gamma-1) + \frac{1}{2}\frac{\partial^2m}{\partial^2\phi}\phi^2 + \frac{1}{2}\frac{\partial^2m}{\partial^2\gamma}(\gamma-1)^2 + \frac{\partial^2m}{\partial\phi\partial\gamma}\phi(\gamma-1)+ \ldots \, .
\end{equation}
Such a mass will lead to a term $m_{,\phi \gamma}(\gamma-1)\phi_{,}^{\phantom{,}j}$ entering the right-hand side of the equations of motion, which in a post-Newtonian expansion would give rise to the presence of the following term:
\begin{equation}
    \frac{\partial^2m}{\partial\phi\partial\gamma}v^2U_s'^j \, ,
\end{equation}
where $U_s\equiv\int\rho^*(1-2s(x'))|\boldsymbol{x-\boldsymbol{x'}}|^{-1}d^3\boldsymbol{x}'$ should be understood as the `scalar-sensitive' Newtonian potential, which in the two-body point-mass limit will lead to $\Sigma_{\gamma\phi}^{(1)}s_{\phi}^{(2)}v^2m_2/r^2$. These considerations lead us to conclude that $\Sigma_{\gamma U}$ ought to be dualled with the $s_U$ of the other body in our equations of motion, as specified above.


For the other sensitivity parameters, in most cases self-consistent definitions of the modified EIH parameters lead to each parameter being dualled only with itself. The only exception to this in the scalar case is with $s_{\Phi_2}$ and $s_U'$, which together make up $\mathcal{D}_{abc}$. However, this is a consequence of the more involved symmetry conditions of  $\mathcal{D}_{122}$ and $\mathcal{D}_{211}$. Another potential exception might be the parameters $s_{\ddot X}$ and $s_\gamma C^-$, which appear in a combination such that the EIH parameters do not prevent their mutual dualling. For them to appear together, however, would require the existence of terms such as $s_{\ddot X}W^j$ and $s_\gamma C^-\ddot X_{,}^{\phantom{,}j}$ in the equations of motion, which seems unlikely given that both potentials are post-Newtonian and enter the equation of motion in isolation.

With these considerations in hand, we may proceed to define the EIH parameters for two sensitive bodies. Rescaling the mass of body $b$ by $m_b \rightarrow m_b/(1-s_\gamma^{(b)})$, we have
\begin{align}
\label{eq:newparams}
    \mathcal{S}_{ab}=&\frac{1}{\left(1-s_\gamma^{(a)}\right)\left(1-s_\gamma^{(b)}\right)}
\, , \qquad
    \mathcal{G}_{ab} = \mathcal{S}_{ab}\left(\alpha-s^{\{ab\}}_N \right) \, ,\\\nonumber
    \mathcal{A}_a=&\frac{a^{(a)}_{s\gamma}}{1-s^{(a)}_\gamma}
    \, , \qquad \qquad
    \mathcal{B}_{ab} = \frac{1}{3}\mathcal{S}_{ab}\left(1-s_\gamma^{(a)}\right)\left(2\gamma+\alpha+\tilde\Sigma_{\gamma U}^{\{ab\}}\right) \, ,\\\nonumber
    \mathcal{C}_{ab}=& \mathcal{S}_{ab}\Big( \alpha_1-\alpha_2 -D^{\{ab\}}+(2\gamma+\alpha)\left(s_\gamma^{(a)}+s_\gamma^{(b)} \right)\\\nonumber
     &\qquad \qquad \qquad -\left(1-s_\gamma^{(a)}\right)\tilde\Sigma_{\gamma U}^{\{ab\}}-\left(1-s_\gamma^{(b)} \right)\tilde\Sigma_{\gamma U}^{\{ba\}}+2s^{\{ab\}}_{\ddot X}+s^{\{ab\}}_N\Big) \, ,\\\nonumber
    \mathcal{D}_{abb}=&\mathcal{S}_{ab}^2 \left( 1-s^{(a)}_\gamma \right)\left(2\beta-\alpha^2-\alpha s^{\{ab\}}_N+s^{\{ab\}}_{{\rm PN}}\right)\, ,\\\nonumber
    \mathcal{E}_{ab}=& \mathcal{S}_{ab}\left(\alpha_2-E^{\{ab\}} -2s^{\{ab\}}_{\ddot X} +s^{\{ab\}}_{U} \right) \, ,
\end{align}
where
\begin{align}
    s_{\rm N}^{\{ab\}} \equiv& s_U^{(a)}+s_U^{(b)}-\frac{1}{c_{\rm N}} s_U^{(a)} s_U^{(b)}
    \, , \qquad
    s^{\{ab\}}_{{\rm PN}} \equiv  s^{(a)\prime}_U +s^{(b)}_{\varphi_2}+ \frac{1}{c_{\rm PN}}  s^{(a)\prime}_U s^{(b)}_{\varphi_2} \, ,\\\nonumber
    \tilde\Sigma^{(a)}_{\gamma U}\equiv& s^{(a)}_U-\Sigma_{\gamma U}^{(a)} \,, \hspace{2.8cm}
    \tilde\Sigma^{\{ab\}}_{\gamma U}\equiv \tilde\Sigma^{(a)}_{\gamma U} +s_U^{(b)}+\frac{1}{c_{N}}\tilde\Sigma^{(a)}_{\gamma U}s_U^{(b)} \, ,\\\nonumber
    E^{(a)}\equiv&2C^-s_\gamma^{(a)} \,,\hspace{3.2cm}
    E^{\{ab\}}\equiv E^{(a)}+E^{(b)}+\frac{1}{c_{E}}E^{(a)}E^{(b)}\, ,\\\nonumber
    D^{(a)}\equiv&(C^+-2C^-)s_\gamma^{(a)} \, , \hspace{1.8cm}
    D^{\{ab\}}\equiv D^{(a)}+D^{(b)}+\frac{1}{c_{D}}D^{(a)}D^{(b)}\, .
\end{align}
These theory-independent parameters can accommodate both scalar-tensor theories and vector-tensor theories, which have had their equations of motion of sensitive compact bodies calculated in References \cite{mirshekari2013compact,taherasghari2023compact,taherasghari2025compact}. However, to our knowledge, no such calculation has been performed for a theory with sensitivities to both scalar and vector degrees of freedom. In the following section, we consider such a theory, and calculate the equation of motion up to first post-Newtonian order, to demonstrate the applicability of our new formalism.

\section{Example: A Scalar-Vector-Tensor Theory of Gravity}
\label{sc:FullTheory}

There are many ways to combine fundamental scalar and vector degrees of freedom in a gravitational theory at the level of the action. We chose to construct a canonical Einstein-\AE ther term in the Einstein conformal frame to ensure that, even in the absence of sensitivities, our theory is only semi-conservative.

\subsection{Definition via an Action}

The Einstein frame is defined  by the metric $g_{\mu\nu} \equiv \varphi \, \tilde g_{\mu\nu}$, where $
\tilde g_{\mu\nu}$ is the metric of the Jordan frame and $\varphi = \phi/\phi_0$, such that
\begin{equation}
\begin{aligned}
\label{eq:SVTaction}
S \equiv {} & \frac{\phi_0}{16\pi} \int d^4x \, \sqrt{-g} \Big[
\ R - E^{\mu\nu}{}_{\alpha\beta} \nabla_\mu K^{\alpha}{}_{ \ae} \nabla_\nu K^{\beta}{}_{ \ae}
+ \lambda \left( K_{ \ae\,\mu} K^{\mu}{}_{ \ae} + 1 \right)
\\
&  \hspace{5cm} - \frac{2\omega(\phi)+3}{\phi^2} g^{\mu\nu} \partial_{\mu}\phi \, \partial_{\nu}\phi \Big] + \int d^4x \, \sqrt{-g} \, \mathcal{L}_M \, ,
\end{aligned}
\end{equation}
where $R$ is the Ricci scalar of the Einstein frame metric $g_{\mu\nu}$, $\nabla_{\mu}$ is its associated covariant derivative, $\phi$ is a scalar field, $\omega(\phi)$ is its coupling function, $K_{{\rm \ae}}^{\mu}$ is the \ae ther field, $\lambda$ is a Lagrange multiplier that is used to enforce the constraint $K_{{\rm \ae}}^{\mu}K_{{\rm \ae} \, \mu}=-1$, and 
\begin{align}
E^{\mu\nu}{}_{\alpha\beta}
\equiv {} & c_1\, g^{\mu\nu} g_{\alpha\beta}
+ c_2\, \delta^\mu_{\alpha} \delta^\nu_{\beta}
+ c_3\, \delta^\mu_{\beta} \delta^\nu_{\alpha}
- c_4\, K^{\mu}{}_{ \ae} K^{\nu}{}_{ \ae} g_{\alpha\beta} \, .
\end{align}
The coupling constants $c_1,c_2,c_3$ and $c_4$ are free parameters of the vector sector of this theory. Following the constraint on gravitational wave speed, provided by GW170817 \cite{GW170817Speed}, we impose the condition that $c_3=-c_1$, which results in this equalling that of light, which remains the case in both Einstein and Jordan conformal frames.  

Varying Eq. (\ref{eq:SVTaction}) with respect to the metric yields 
\begin{align}
G_{\mu\nu}
- \frac{2\omega + 3}{2\phi^{2}}
\left(
\partial_{\mu}\phi \, \partial_{\nu}\phi
- \frac{1}{2}
g_{\mu\nu}
g^{\alpha\beta}
\partial_{\alpha}\phi \, \partial_{\beta}\phi
\right)
= \frac{8\pi}{\phi_0}  \left(T_{\mu\nu}- K^{\mu}_{ \ae} K^{\nu}_{ \ae}
K_{ \ae\,\alpha} T^{\alpha}_{ \ae}\right)+S_{\mu\nu}\, ,
\end{align}
where we note that $T_{\mu\nu}$ is related to the energy-momentum tensor in the Jordan frame by $T_{\mu\nu}= (\phi_0/\phi)\tilde T_{\mu\nu}$. Varying Equation (\ref{eq:SVTaction}) with respect to $K_{\ae}^{\mu}$ yields 
\begin{align}
\label{eq:veq}
\hspace{-0.5cm}
c_1 \nabla_\nu F^{\mu\nu}
&= 8\pi T^{\mu}_{ \ae}
- \lambda K^{\mu}_{ \ae}
- c_2 \nabla^\mu \left( \nabla_\nu K^{\nu}_{ \ae} \right)
- c_4 \left(
a^{\nu}_{ \ae} \nabla^\mu K_{ \ae\,\nu}
- a^{\mu}_{ \ae} \nabla_\nu K^{\nu}_{ \ae}
- K^{\nu}_{ \ae} \nabla_\nu a^{\mu}_{ \ae}
\right),
\end{align}
where $F_{\mu\nu} \equiv \partial_\nu K_{ \ae\,\mu}
- \partial_\mu K_{ \ae\,\nu}$ and $a^{\mu}_{ \ae} \equiv K^{\nu}_{ \ae} \nabla_\nu K^{\mu}_{ \ae}$, and varying with respect to $\phi$ gives
\begin{align}
g^{\mu\nu}\nabla_{\mu}\nabla_{\nu}\phi
+ \frac{1}{2}\frac{d}{d\phi}
\left[
\ln\!\left(\frac{2\omega + 3}{\phi^{2}}\right)
\right]
g^{\mu\nu}
\partial_{\mu}\phi \, \partial_{\nu}\phi
=
\frac{8\pi}{\phi_0}
\frac{\phi}{2\omega + 3}
T\, ,
\end{align} 
where $T\equiv g^{\mu\nu}T_{\mu\nu}$. Requiring the vector field to be unit time-like leads to the additional constraint
\begin{align}
\lambda
&=
-8\pi K_{ \ae\,\nu} T^{\nu}_{ \ae}
- \frac{1}{2} c_1
\left(
F_{\mu\nu} F^{\mu\nu}
+ 2 \nabla_\nu a^{\nu}_{ \ae}
\right) 
+ c_2 K^{\nu}_{ \ae} \nabla_\nu \left( \nabla_\mu K^{\mu}_{ \ae} \right)
+ 2 c_4 \left| a_{ \ae} \right|^2 \,.
\end{align}
Here we define the energy-momentum tensor and vector by
\begin{align}
T^{\mu\nu}
&\equiv
\frac{2}{\sqrt{-g}}
\frac{\delta \left( \sqrt{-g}\,\mathcal{L}_M \right)}{\delta g_{\mu\nu}} \qquad\text{and}\qquad
T_{ \ae\,\mu}
\equiv
- \frac{1}{\sqrt{-g}}
\frac{\delta \left( \sqrt{-g}\,\mathcal{L}_M \right)}{\delta K^{\mu}_{ \ae}} \,,
\end{align}
and have additionally defined
\begin{equation}
\begin{aligned}
S^{\mu\nu}
&\equiv c_1 \Big[
2 K^{(\mu}_{ \ae} \nabla_\alpha F^{\nu)\alpha}
+ F^{\mu}{}_{\alpha} F^{\alpha\nu}
- K^{\mu}_{ \ae} K^{\nu}_{ \ae}
  (\nabla_\alpha a^{\alpha}_{ \ae}) 
- \frac{1}{4}
\left(
g^{\mu\nu}
+ 2 K^{\mu}_{ \ae} K^{\nu}_{ \ae}
\right)
F_{\alpha\beta} F^{\alpha\beta}
\Big]
\\[6pt]
&\quad
+ c_2 \Big[
\left(
g^{\mu\nu}
+ K^{\mu}_{ \ae} K^{\nu}_{ \ae}
\right)
K^{\beta}_{ \ae}
\nabla_\beta (\nabla_\alpha K^{\alpha}_{ \ae})
+ \frac{1}{2}
g^{\mu\nu}
(\nabla_\alpha K^{\alpha}_{ \ae})^2
\Big]
\\[6pt]
&\quad
- c_4 \Big[
a^{\mu}_{ \ae} a^{\nu}_{ \ae}
- \frac{1}{2}
\left(
g^{\mu\nu}
+ 4 K^{\mu}_{ \ae} K^{\nu}_{ \ae}
\right)
|a_{ \ae}|^2
- K^{\mu}_{ \ae} K^{\nu}_{ \ae}
(\nabla_\alpha a^{\alpha}_{ \ae})
\\
&\hspace{3cm}
- 2 a^{\alpha}_{ \ae}
(\nabla_\alpha K^{(\mu}_{ \ae}) K^{\nu)}_{ \ae}
+ 2 a^{(\mu}_{ \ae}
K^{\nu)}_{ \ae}
\nabla_\alpha K^{\alpha}_{ \ae}
+ 2 K^{\alpha}_{ \ae}
(\nabla_\alpha a^{(\mu}_{ \ae})
K^{\nu)}_{ \ae}
\Big] \,.
\end{aligned}
\end{equation}
This specifies our theory, which we can now analyse in the post-Newtonian limit.

\subsection{Relaxed Field Equations}

To obtain the equation of motion for compact bodies, we must solve the field equations in the near zone.
To do so, we recast the equations into their relaxed form by defining the following: 
\begin{align}
\mathfrak{g}^{\mu\nu} &\equiv \sqrt{-g} \, g^{\mu\nu}  \qquad {\rm and} \qquad
H^{\mu\alpha\nu\beta} \equiv g^{\mu\nu} g^{\alpha\beta}
- g^{\alpha\nu} g^{\beta\mu} \, .
\end{align}
For any space-time we can then write
\begin{align}
\label{eq:LL}
H^{\mu\alpha\nu\beta}{}_{,\alpha\beta}
= (-g)\left(2G^{\mu\nu} + 16\pi \, t^{\mu\nu}_{LL}\right) \, ,
\end{align}
where $t^{\mu\nu}_{LL}$ is the Landau-Lifshitz pseudo-tensor (see Eq. (6.5) of Ref. \cite{poisson2014gravity}). To proceed, we assume that far away from an isolated source the metric is that of Minkowski space, $\eta_{\mu\nu}$, the scalar field tends to a constant asymptotic value, $\phi_0$, and that the system is at rest with respect to the rest frame of the Universe, as set by the asymptotic value of $K^{\mu}_{\ae}$. This requires the leading-order parts of the spatial components of $K^{\mu}_{\ae}$ to be $\sim v^3/c^3$. We can now implicitly define the gravitational potentials, $ h^{\mu\nu}$, by ${\mathfrak{g}}^{\mu\nu} \equiv \eta^{\mu\nu}-  h^{\mu\nu}$, and can impose the  harmonic gauge condition, $\partial_\nu  h^{\mu\nu} = 0$, to get
\begin{align}
\Box_{\eta}\, h^{\mu\nu}
=
-16\pi\,\tau^{\mu\nu}
\qquad {\rm and} \qquad
\Box_{\eta}\varphi = -8\pi \tau_{s} \, ,
\end{align}
where $\Box_{\eta}$ is the d'Alembertian with respect to $\eta_{\mu\nu}$ and where
\begin{align} \nonumber
\tau^{\mu\nu}
\equiv&
(-g)\,\frac{1}{\phi_0}\,(T^{\mu\nu}-K^{\mu}_{ \ae} K^{\nu}_{ \ae}
K_{ \ae\,\alpha} T^{\alpha}_{ \ae})+
 \frac{1}{8\pi}(- g)  S^{\mu\nu} 
+ \frac{1}{16\pi} \left(\Lambda^{\mu\nu}
+ \Lambda^{\mu\nu}_{S} \right)
\,,\\ \nonumber
\tau_{s} \equiv&
-\frac{\sqrt{-g}}{3+2\omega}\frac{\varphi}{\phi_{0}}
\left(
T - 2\varphi \frac{\partial T}{\partial \varphi}
\right)
-\frac{h^{\alpha\beta}\varphi_{,\alpha\beta}}{8\pi}
+
\frac{g^{\alpha\beta}\varphi_{,\alpha}\varphi_{,\beta}}{16\pi}\frac{d}{d\varphi}
\left[
\ln\!\left(\frac{3+2\omega}{\varphi^{2}}\right)
\right]
 ,\\[-1cm] \nonumber
\end{align}
and
\vspace{-0.1cm}
\begin{align}
\nonumber\hspace{-3cm}
\Lambda^{\mu\nu}\equiv& \,
16\pi
(- g)\,  t^{\mu\nu}_{LL}
+
 h^{\mu\alpha}{}_{,\beta}
\, h^{\nu\beta}{}_{,\alpha}
-
 h^{\alpha\beta}
\, h^{\mu\nu}{}_{,\alpha\beta}\,,
\\ \nonumber
 \hspace{-3cm}\Lambda^{\mu\nu}_{S}\equiv&\,
2(- g)\frac{3+2\omega}{\varphi^{2}}
\,\varphi_{,\alpha}\,\varphi_{,\beta}
\left(
\mathfrak{g}^{\mu\alpha}\mathfrak{g}^{\nu\beta}
-\frac{1}{2}\mathfrak{g}^{\mu\nu}\mathfrak{g}^{\alpha\beta}
\right) \, .
\end{align}
We can now proceed to solve these equations, by iterating these equations in terms of the smallness parameter $\epsilon \sim v^2 \sim m/r$.

\subsection{Metric in the Near Zone}


We can now define the following notation for the components of the field potential $ h_{\mu\nu}$:
\begin{align} \nonumber
N &\equiv h^{00} \sim \epsilon , \quad K^{j} \equiv h^{0j} \sim \epsilon^{3/2}, \quad B^{jk} \equiv h^{jk} \sim \epsilon^{2}, \\ \nonumber
B &\equiv h^{jj}
\equiv \sum_j h^{jj}
\sim \epsilon^{2} , \quad \Psi \equiv \varphi - 1 \sim \epsilon \,,
\end{align}
%
so that the vacuum equations are de-sourced by the following transformations \cite{taherasghari2023compact}:
\begin{align} \nonumber
N &\;\rightarrow\;N
+ \frac{\epsilon c_{14}}{2-c_{14}} B
+ \frac{\epsilon}{v_L^2}\dot R \,, \qquad K^j \;\rightarrow\; K^{j}
- R_{,j}
\\[6pt] \nonumber
B^{jk} &\;\rightarrow\;  B^{jk}
+ \delta^{jk}
\dot R \, , \qquad K_{ \ae}^{\,j}
\;\rightarrow\;
K^{j}_{\ae}
+ K^{j}
+ \frac{1}{2c_{14}}
\left(
W_L R_{,j}
+ c_1 W_T X_{K\ae ,j}
\right) ,
\end{align}
where $v_T^2 \equiv {c_1}/{c_{14}}$, $v_L^2 \equiv {c_2(2 - c_{14})}/{c_{14}(2 + 3c_2)}$, $W_T \equiv 1 - {v_T^{-2}}$ and $W_L \equiv (1 - \frac{1}{2}c_{14})(1 - {v_L^{-2}})$, and where we have defined the super-potentials $\nabla^2 X_N \equiv 2N$ and $\nabla^2 X_{K\, \ae} \equiv 2 K^{k}_{ \ae,\,k}$ such that $R = \frac{1}{4} c_{14}\dot{X}_N - c_1 X_{K \ae}$ (see Ref. \cite{taherasghari2023compact} for further details of this approach).
%
%

To the required order, the de-sourced field equations now read
\begin{equation} 
\label{eq:desourcedFE}
\begin{aligned}
    \Box\left(\left(1-\frac{c_{14}}{2} \right)N \right)&=
    -16\pi  \tau^{00}, \qquad
\Box K^{j*}
= -16\pi  \tau^{0j} ,
\\[6pt]
\Box B^{jk}
&= -16\pi  \tau^{jk}
,
\qquad
\Box B
= -16\pi  \tau^{kk}
, \\[-0.25cm] 
\end{aligned}
\end{equation}
and
\begin{equation} \label{dsd2}
\phantom{a} \hspace{0.75cm} c_1 \Box^{*} K_{ \ae}^{\,j}
= 8\pi \tau_{ \ae}^{\,j}
 , \hspace{1.5cm}
\Box\,\Psi
=
-8\pi\,\tau_s ,
\end{equation}
where
\(
\Box^{*} \equiv \nabla^{2} - v_T^{2}\partial_{0}^{2}
\) and \(c_{14}\equiv c_1+c_4\). The leading-order parts of the metric potentials are then given by
\begin{equation} \nonumber
N
=
\frac{4}{\alpha} \int \frac{\tau^{00}(t,\mathbf{x}')}{|\mathbf{x}-\mathbf{x}'|}\, d^3x'
\qquad
{\rm and}
\qquad
\Psi
=
2 \int \frac{\tau_s(t,\mathbf{x}')}{|\mathbf{x}-\mathbf{x}'|}
\, d^3x'
\,,
\end{equation}
where we have defined $\alpha \equiv \left(1-\frac{1}{2}c_{14}\right)$. We can now define the pseudo-tensors to the required order, noting that $\Lambda^{\mu\nu}$ has been combined with the contribution from $ S^{\mu\nu}$ such that $\Lambda^{\mu\nu} \rightarrow \Lambda^{\mu\nu}+S^{\mu\nu}$, 
\begin{align} \nonumber
&\Lambda^{00}= -\frac{7}{16}(2 - c_{14})N_{,j}^{\,2} ,\qquad \Lambda^{jj}=-\frac{1}{16}(2-c_{14})N_{,j}^{\,2} 
,\qquad \Lambda^{00}_s=\frac{3+2\omega_0}{2} \nabla \Psi_{,\,j}^2\,,
\end{align}
to get
\begin{equation}
\begin{aligned}
\tau^{00} &=  G(1-\zeta)
\left\{
\sigma - \sigma^{ii}
+ G(1-\zeta)\!\left(4\sigma U - \frac{7}{8\pi}(\nabla U)^2\right)
- G\zeta\!\left(6\sigma U_s - \frac{1}{8\pi}(\nabla U_s)^2\right)
\right\}
\\[6pt]
\tau^{ii} &= \alpha G(1-\zeta)
\left\{
\sigma^{ii}
- \frac{1}{8\pi}G(1-\zeta)(\nabla U)^2
- \frac{1}{8\pi}G\zeta(\nabla U_s)^2
\right\}
\\[6pt]
\tau_s &=  G\zeta
\left\{
\sigma_s
+ 2G(1-\zeta)\sigma_s U
- 2G(2\lambda_1+\zeta)\sigma_s U_s
+ \frac{1}{2\pi}G(\lambda_1-\zeta)(\nabla U_s)^2
\right\}
\\[10pt]
\tau^{0i} &= \alpha G(1-\zeta)\sigma^{i}
, \qquad
\tau^{ij} = 0
, \qquad
\tau^{j}_{\ae} = \alpha G(1-\zeta)\sigma^{j}_{\ae} 
\, ,
\end{aligned}
\end{equation}
%
%
where we have re-scaled $\Psi\rightarrow\alpha\Psi$, and where we have defined
\begin{align} \nonumber
G \equiv \frac{1}{\alpha\phi_0}\,\frac{(4 + 2\omega_0)}{(3 + 2\omega_0)}, \qquad
\zeta \equiv \frac{1}{(4 + 2\omega_0)}, \qquad
\lambda_1 \equiv
\zeta\frac{\left({d\omega}/{d\varphi}\right)_0 }{(3 + 2\omega_0)},
\end{align}
and used the $\sigma$ densities \cite{blanchet1989post}
\begin{align} \nonumber
\sigma &\equiv \frac{1}{\varphi} (T^{00} + T^{ii}), \quad
\sigma^i \equiv \frac{1}{\varphi} T^{0i}, \quad
\sigma^{ij} \equiv \frac{1}{\varphi}T^{ij}, \quad
\sigma^{j}_{ \ae} \equiv T^{j}_{ \ae} \,\quad
\sigma_s \equiv  -T + 2\varphi \frac{\partial T}{\partial \varphi}.
\end{align}
We can now insert the above expressions into the de-sourced field equations (\ref{eq:desourcedFE})-(\ref{dsd2}) to obtain the following, where the definitions of the potentials are given in Appendix C:
\begin{equation}
\begin{aligned}
N &=  {4 (1-\zeta) G U} + G(1-\zeta)
\Big\{
7G(1-\zeta)U^2
- 4\Phi_1
+ 2G(1-\zeta)\Phi_2
\\&\hspace{6cm}+ 2\ddot X
- G\zeta U_s^2
- 24G\zeta \Phi_{2s}
+ 2G\zeta \Phi^{s}_{2s}
\Big\},
\\[0pt]
B &= \alpha G(1-\zeta)
\Big\{
G(1-\zeta)U^2
+ 4\Phi_1
- 2G(1-\zeta)\Phi_2
+ G\zeta U_s^2
- 2G\zeta \Phi^{s}_{2s}
\Big\},
\\[6pt]
\Psi &= {2 \alpha G \zeta U_s} + 2G\zeta
\Big\{
-2G(\lambda_1-\zeta)U_s^2
+ 4G(1-\zeta)\Phi^{s}_{2}
- 4G(\lambda_1+2\zeta)\Phi^{s}_{2s}
+ \ddot X_s
\Big\},
\\[6pt]
K^{i} &= 4\alpha G(1-\zeta)V^{i}, \qquad 
R
= c_{14}\,G(1-\zeta)\,\dot X
+ 2\,G(1-\zeta)\,\alpha\,X^{j}_{\ae,j},
\\[6pt]
\end{aligned}
\end{equation}
with
\begin{equation}
\\
K^{j}_{\ae}
= -2\,\alpha\,G(1-\zeta)\,c_1^{-1} V^{j}_{\ae} \qquad {\rm and} \qquad
X_{K \ae}
= -\frac{2G(1-\zeta)}{c_1}\,\alpha\,X^{j}_{\ae,j}.
\end{equation}
Undoing the transformations above, and applying the gauge transformation $t \rightarrow t+ \xi^0$ with 
\begin{equation}
\begin{aligned}
\xi_0
&=
\frac{1}{4}c_{14}(3+v_L^{-2})
G(1-\zeta)\,\dot X
+ \frac{1}{2}\,\alpha\,(3+v_L^{-2})\,G(1-\zeta)\,X^{k}_{\ae,k} \, ,
\end{aligned}
\end{equation}
we obtain the following metric to 1PN:
\begin{equation} \label{gst}
\begin{aligned}
g_{00} ={}&
-1
+ 2G(1-\zeta)U
+ 2 G\zeta U_s
- 2G^2(1-\zeta)^2 U^2
- 2G^2\zeta(\zeta+\lambda_1)U_s^2
\\[4pt]
&- 4 G^2\zeta(1-\zeta)U U_s
+ 4 G^2\zeta(1-\zeta)\Phi^{s}_{2}
- 12 G^2\zeta(1-\zeta)\Phi_{2s}
\\[4pt]
&- 4\alpha^2G^2\zeta(2\zeta+\lambda_1)\Phi^{s}_{2s}
+ G(1-\zeta)\ddot X
+ G\zeta \ddot X_s \\
g_{0j} ={}&
-4\alpha\,G(1-\zeta)\,V^{j}
+c_{14}\frac{1}{2}\,(1-v_L^{-2})\,G(1-\zeta)\,\dot X_{,j}
+\frac{1}{2}\alpha\,(1-v_L^{-2})\,G(1-\zeta)\,X^{k}_{\ae,jk}
\\[4pt]
g_{ij}
={}&\delta_{ij}
\left[
1 + 2G(1-\zeta)U - 2\alpha G\zeta U_s
\right],
\end{aligned}
\end{equation}
with the spatial part of the \ae ther field, to the required order, as
\begin{align}
K^j_{\ae}&= -2\alpha\,G(1-\zeta)\,c_1^{-1} V^{j}_{\ae}
+ 4\alpha\,G(1-\zeta)\,V^{j} 
\\[4pt] 
&\qquad \qquad
+ \frac{1}{2} W_L\,G(1-\zeta)\,\dot{X}_{,j}
+ \frac{\alpha}{c_{14}}(W_L - W_T)\,G(1-\zeta)\,X^{k}_{\ae,jk}. \nonumber
\end{align}
This result can be used to calculate the equation of motion of a body in this theory.

\subsection{Equations of Motion for Compact Bodies}

We now want to calculate the motion of compact bodies within this theory, by allowing the matter source to be sensitive to both the scalar and vector fields, $\phi$ and $K^{\mu}_{\ae}$. We remind the reader that whilst the metric potentials were calculated in the Einstein frame, the metric itself is written for the Jordan frame, where we are now working. For the scalar field we follow Eardley and allow the mass to depend explicitly on the scalar field, such that $m=m(\phi)$ \cite{eardley1975observable}. For the vector field we follow Reference \cite{taherasghari2023compact} and write $m=m(\gamma)$, where $\gamma\equiv -\varphi^{1/2} K_{\ae\,\mu}u^{\mu}$ and $u^\mu$ is the four-velocity of the body (the factor of $\varphi^{1/2}$ is introduced so that the vector field is normalised in the Jordan frame). This gives the mass an overall dependence $m=m(\phi,\gamma)$, so that after a post-Newtonian expansion
\begin{equation}
    m(\phi,\gamma)= m_0+\frac{\partial m}{\partial\phi}\phi+\frac{\partial m}{\partial \gamma}(\gamma-1) + \frac{\partial^2m}{\partial^2\phi}\phi^2 + \frac{\partial^2m}{\partial^2\gamma}(\gamma-1)^2 + \frac{\partial^2m}{\partial\phi\partial\gamma}\phi(\gamma-1)+ \ldots .
\end{equation}
The matter action is then constructed from this mass via
\begin{equation}
S_m = - \sum_A \int m_A(\phi,\gamma)\, d\tau_A,
\end{equation}
where $\tau_A$ is the proper time along the worldline of the $A$-th body. By varying with respect to the metric and the \ae ther field, we obtain the following expressions for the energy-momentum tensor and vector, respectively:
\begin{equation}
\begin{aligned}
T^{\mu\nu}
&=
\sum_A
\frac{\delta^{3}(x-x_A)}{u^{0}\sqrt{-g}}
\Bigg[
\left(
m_A(\phi,\gamma)
-\gamma\,\frac{\partial m_A(\phi,\gamma)}{\partial \gamma}
\right)
u^{\mu}u^{\nu}
+2\,\frac{\partial m_A(\phi,\gamma)}{\partial \gamma}
\,u^{(\mu}K^{\nu)}_{\ae}
\Bigg],
\\[10pt]
T_{\ae}^{\mu}
&=
-\sum_A
\frac{\delta^{3}(x-x_A)}{u^{0}\sqrt{-g}}
\frac{\partial m_A(\phi,\gamma)}{\partial \gamma}
\,u^{\mu}.
\end{aligned}
\end{equation}
We can now define the necessary sensitivity parameters to 1PN order as follows: 
\begin{equation}
\begin{aligned}
s_{\phi}
&\equiv
\frac{\partial \ln m(\phi,\gamma)}
{\partial \ln \phi}
,
\qquad
s_{\gamma}
\equiv
\frac{\partial \ln m(\phi,\gamma)}
{\partial \ln \gamma}
,
\qquad 
s_{\gamma\phi}
\equiv
\frac{\partial^2 \ln m(\phi,\gamma)}
{\partial  \ln\gamma \partial \ln\phi}
,
\\[10pt]
s'_{\phi}
&\equiv
\frac{\partial^{2}\ln m(\phi,\gamma)}
{\partial(\ln \phi)^{2}}
,
\qquad
s'_{\gamma}
\equiv
\frac{\partial^{2}\ln m(\phi,\gamma)}
{\partial(\ln \gamma)^{2}}
\,,
\end{aligned}
\end{equation}
where each of these should be understood to be evaluated at $\phi=\phi_0$ and $\gamma=1$. It is also convenient to define $a_\gamma \equiv s_\gamma^{(2)}-s_\gamma+s_\gamma'$ and $a_\phi \equiv s_\phi^2-\frac{1}{2}s_\phi+s_\phi'$.
Using these, and treating the energy density $\rho^{*}$ as corresponding to that of a set of point masses, we can write the sigma densities in the presence of sensitive bodies as
\begin{align}
\sigma
&= \rho^{*}
\left[
1
+ \epsilon\left\{
\frac{3}{2}(1-s_\gamma)v^{2}
- G(1-\zeta)U_{\sigma}
+ \alpha G\zeta(5+2s_{\phi})U_{s\sigma}
\right\}
\right] , \\
\sigma_s
&\nonumber=
\rho^{*}
\left[
(1-2s_{\phi})
-\epsilon
\left\{
\frac{1}{2}(1-s_{\gamma}-2s_{\phi}+2s_{\gamma\phi}-2s_\gamma s_\phi)v^{2}
\right. \right.\\
& \qquad \quad + 3G(1-\zeta)(1-2s)U_{\sigma}  - \left. \left. 3\alpha G\zeta\left(1-2s_{\phi}-\frac{4}{3}a_{\phi}\right)U_{s\sigma}
\right\}
\right] ,
\\[6pt]
\sigma^{i}
&= \rho^{*} (1-s_\gamma)v^{i} \, , \qquad
\sigma^{ij}
= \rho^{*} v^{i} v^{j}(1-s_\gamma) \, , \qquad
\sigma_{\ae}^i=-s_\gamma\rho^{*}v^i .
\end{align}
Substituting these into Equation (\ref{gst}) gives the metric in the presence of these bodies, which we can now use to calculate
%
the equation of motion of such bodies to first post-Newtonian accuracy. 


From the equation of motion (\ref{eq:geo2}), the Newtonian-order part of this is given by
\begin{align}
a^{i}_{N}
= \frac{
G(1-\zeta)\,U^{,i}
+ G\zeta(1-2s_\phi)\,U_{s}^{,i}
}{
1 - s_{\gamma}
} \, , 
\end{align}
with the post-Newtonian contribution being given by
\begin{align}
a^{i}_{\mathrm{PN}} 
&= \Bigg[3GU^{,i}(1-\zeta)
- G\zeta\!\left(1 + \frac{2s_{\gamma\phi}}{1-s_{\gamma}} - 2s_{\phi}\right) U^{,i}_{s}
\nonumber 
- \frac{\left(1 + \frac{a_{\gamma}}{1-s_{\gamma}}\right)
\left[\,G U^{,i} (1-\zeta) + G\zeta(1-2s_{\phi}) U^{,i}_{s}\,\right]}
{1-s_{\gamma}} \Bigg]\frac{v^2}{2}\\
&-\Bigg[3GU^{,j}(1-\zeta)
- G\zeta\!\left(1 + \frac{2s_{\gamma\phi}}{1-s_{\gamma}} - 2s_{\phi}\right) U^{,j}_{s}
\nonumber 
+\frac{\left(1 + \frac{a_{\gamma}}{1-s_{\gamma}}\right)
\left[\,G U^{,j} (1-\zeta) + G\zeta(1-2s_{\phi}) U^{,j}_{s}\,\right]}
{1-s_{\gamma}} \Bigg] v^jv^i
  \nonumber \\[4pt]
&\quad
- v^{i}\Big[
3G(1-\zeta)\dot U
- G\zeta \left(1-2s_{\phi}+\frac{2s_{\gamma\phi}}{1-s_\gamma} \right)\dot U_{s}
\Big]  \nonumber \\[4pt]
&\quad
+\frac{1}{1-s_\gamma}\Bigg[-4G^{2}(1-\zeta)^{2}UU^{,i}
- 4G^{2}\zeta(1-\zeta)\left(1-2s_{\phi}+\frac{2s_{\gamma\phi}}{1-s_\gamma} \right)UU_{s}^{,i}
  \nonumber \\[4pt]
&\qquad \qquad
- 2G^{2}\zeta\!\left[\left(\lambda_{1}+\frac{2s_{\gamma\phi}}{1-s_\gamma} \right)(1-2s_{\phi})+2\zeta s'_{\phi}\right] U_{s}U_{s}^{,i}
+ \tfrac12 G(1-\zeta)\ddot X^{,i}
+ \tfrac12 G\zeta(1-2s_{\phi})\ddot X_{s}^{,i}  \nonumber \\[4pt]
&\qquad \qquad
+ \tfrac32 G(1-\zeta)\Phi_{1}^{,i}
- \tfrac12 G\zeta(1-2s_{\phi})\Phi_{1}^{s,i}
- G^{2}(1-\zeta)^{2}\Phi_{2}^{,i}
- G^{2}\zeta(1-\zeta)(1-2s_{\phi})\Phi_{2}^{s,i}  \nonumber \\[4pt]
&\qquad \qquad
- G^{2}\zeta\!\left[1-\zeta+(2\lambda_{1}+\zeta)(1-2s_{\phi})\right]\Phi_{2s_{\phi}}^{s,i}
- 4G^{2}\zeta^{2}(1-2s_{\phi})\Sigma^{i}(a_{s}U_{s}) \Bigg].\nonumber\\
&+ 8\!\left(1-\frac{c_{14}}{2}\right) G(1-\zeta)\, V^{[i,j]} v^{j}
- \frac{(2-c_{14})s_{\gamma}}{(1-s_{\gamma})c_{1}}
\,G(1-\zeta)\,\dot V^{i}_{\ae}
\nonumber\\[6pt]
&\quad
- \frac{2(2-c_{14})s_{\gamma}}{c_{1}(1-s_{\gamma})}
\,G(1-\zeta)\, V^{[i,j]}_{\ae} v^{j}
\nonumber
+\frac{1}{4}\!\left(
\frac{s_{\gamma}(2-c_{14})+c_{14}}{(1-s_{\gamma})v_{L}^{2}}
\right)
G(1-\zeta)\,\ddot X^{,i}
\nonumber\\[6pt]
&\quad
+ \frac{1}{2(1-s_{\gamma})}
\left(
W_{L}
+\frac{(2-c_{14})(W_{L}-W_{T})s_{\gamma}}{c_{14}}
\right)
G(1-\zeta)\, \dot X^{j,i}_{\ae,j}+\, 4\!\left(1 - \frac{c_{14}}{2}\right)
G(1-\zeta)\,\dot V^{i}.
\end{align}
The EIH parameters for the two-body equations of motion for our theory can then be read off as follows, where the masses have been rescaled as $m_a\rightarrow m_a/(1-s_\gamma^a)$: 

\begin{align}
\mathcal{G}_{12}
&=\frac{1}{(1-s_\gamma^{(1)})(1-s_\gamma^{(2)})}\bigg[
G (1 - \zeta)
+ G \zeta (1 - 2 s_{\phi}^{(1)})(1 - 2 s_{\phi}^{(2)})\bigg]\\ \nonumber
\mathcal{A}_1
&=\frac{a^{(1)}_\gamma}{1-s^{(1)}_\gamma}, \qquad
\mathcal{B}_{12} 
= \frac{1}{3 \left(1-s_\gamma^{(2)}\right)}\left[3G(1-\zeta)
-G\zeta
\left(1-2s^{(1)}_{\phi}+\frac{2s^{(1)}_{\gamma\phi}}{1-s^{(1)}_{\gamma}}\right)
(1-2s_{\phi}^{(2)})\right]
\\ \nonumber
\mathcal{C}_{12}&=\frac{2 (2-c_{14})G(1-\zeta)}{(1-s_\gamma^{(1)})(1-s_\gamma^{(2)})}\big((1-s_\gamma^{(1)})(1-s_\gamma^{(2)})+\frac{s_\gamma^{(1)}s_\gamma^{(2)}}{c_1}\big)-\mathcal{E}_{12}
-2\mathcal{G}_{12}-6\mathcal{B}_{(12)}
\\
\mathcal{D}_{122}
&
= \frac{1}{(1 - s^{(1)}_{\gamma})(1-s^{(2)}_\gamma)^2}
\Bigg[\bigg(
G (1 - \zeta)
+ G \zeta (1 - 2 s_{\phi}^{(1)})(1 - 2 s_{\phi}^{(2)})\bigg)^{2}
\nonumber \\[-8pt] \nonumber
&\hspace{5cm}
+ 2G \zeta (1 - 2 s^{(2)}_{\phi})^{2}
\left(
\lambda_1(1 - 2 s^{(1)}_{\phi}) + 2 \zeta s'^{(1)}_{\phi}
\right)
\Bigg]
\\ \nonumber
\mathcal{E}_{12}
&=\frac{2 G(1-\zeta)}{(1-s_\gamma^{(1)})(1-s_\gamma^{(2)})}\bigg[\frac{(2 - c_{14})}{2c_{14}} \, s^{(1)}_\gamma s^{(2)}_\gamma \, W_T  - \frac{\left(c_{14} + (2 - c_{14}) s^{(1)}_\gamma\right)\left(c_{14} + (2 - c_{14}) s^{(2)}_\gamma\right)}{4c_{14}} \, W'_L\bigg] ,
\end{align}
where $W'_L=1-v_L^{-2}$. We can now map our theory-independent sensitivity parameters to this theory to find
\begin{align}
s_U^i =& 2G \, \zeta \, s_\phi^i, \quad s_U'^i = 2G^2\zeta(-\lambda_1 s_\phi^i+\zeta s'^{i}_\phi) \,{\rm},
\quad s_{\varphi_2}^i = -4\zeta\lambda_1G^2 \left(s_\phi^i+(s_\phi^i)^2 \right),\\
        s_\gamma^i=&s_{\gamma, \ae}^i , \qquad
        \Sigma^i_{\gamma U}= \frac{2G\zeta s_{\gamma \phi}^i}{(1-s_\gamma^i)},
\end{align}
and with critical values
\begin{equation}
c_N=G\zeta  \qquad {\rm and} \qquad c_{\rm PN}=G^2\zeta\lambda_1 \, .
\end{equation}
The remaining constants are specified by solving
\begin{align} \nonumber
E^i=&\frac{G(1-\zeta)(2-c_{14})W_L'}{2c_{14}}s_\gamma^i\, , \quad
\left(\frac{E^i}{ s_\gamma^i}\right)^2=\frac{G(1-\zeta)(2-c_{14}) c_E}{2c_{14}}\left[(2-c_{14})W_L'-2W_T\right]\\ \nonumber
D^i=&2(2-c_{14})G(1-\zeta)s_\gamma^i-E^i, \; \left(\frac{D^i}{ s_\gamma^i}\right)^2=2c_D(2-c_{14})G(1-\zeta)(1+c_1^{-1})- \frac{c_D}{c_E}\left(\frac{E^i}{s_\gamma^i}\right)^2.
\end{align}
This example explicitly demonstrates our theory-independent formalism's ability to describe complicated theories that are sensitive to both scalars and vectors.

\section{Discussion}
\label{sc:Diss}

The formalism we have developed here provides a basis for including both scalar and vector sensitivities in the equations of motion of compact bodies in theories of gravity that violate the strong equivalence principle. It accommodates known results for sensitive bodies in the existing literature, and extends them to new situations. In addition, it possesses all the degrees of freedom of the modified EIH formalism for the motion of compact bodies, while offering the added benefit of distinguishing between weak and strong-field effects. Nevertheless, despite these successes, it is still possible to go further. Our formalism currently only applies to semi-conservative theories (i.e. those that admit a Lagrangian-based description of dynamics in at least one frame). Theories that do not possess concepts of conserved energy or momentum are currently beyond the scope of what we have done. There are also variations or extensions to our formalism that have not yet been directly considered. Most obviously, the spin of the bodies has not been taken into account. This would change the dynamics directly and would also present a new class of invariants for the mass of a compact body to be sensitive to, i.e. $K^{\mu}S_{\mu}$, where $S^\mu$ and $K^\mu$ represent the spin and vector field, respectively. It should also be possible to allow the mass to depend upon different kinds of invariants, as considered in Reference \cite{taherasghari2022modified}, such as $C^{\mu\nu}u_{\mu}u_{\nu}$ where $C^{\mu\nu}$ is some external gravitational field and $u_{\mu}$ is a body's 4-velocity (one could equally contract with spin instead). Should a theory that possesses these degrees of freedom, or similar ones, emerge, we expect it should be possible to further extend our formalism by re-deriving the theory-independent equation of motion, and then proceeding as we have done here. 

We note that it should also possible to use observational data of gravitational waves from compact binary coalescences to place constraints on the theory-independent parameters describing the motion of the bodies, as was done in Reference \cite{pitt2025constraining}. This was achieved by deriving an expression for the gravitational wave phase as a function of frequency, which involved assuming: (i) a theory-independent expression for the gravitational flux from the system, (ii) an energy-balance argument between the decrease in binding energy and flux of gravitational waves, and (iii) the stationary-phase approximation. To re-derive the phase expression for the present case we do not need to change any of these assumptions, and the procedure would be identical to that described in Reference \cite{pitt2025constraining}, except for a different equation of motion and binding energy. As such, we choose not to re-produce the full gravitational wave phase derivation here. 

\section{Conclusions}
\label{sc:Conc}
We have extended theory-independent formulations of gravity to describe the motion of compact bodies in theories with preferred frames. In the case of frame-independent theories, the motion of compact bodies is characterised by two sets of parameters: (i) the PPN parameters governing weak-field metric effects, and (ii) sensitivities that encode the changes to motion coming from strong field effects inside compact bodies. Our extension to theories with preferred frames has postulated the existence of a post-Newtonian vector field that defines a universal rest frame, and which permits the masses of compact bodies to be sensitive to this field contracted with the bodies' own four-velocity. This assumption introduces a new kind of inertial sensitivity, in addition to new gravitational sensitivities of the kind that had been considered up to this point. We validated this new formalism by constructing a scalar-vector-tensor theory and calculating the motion of compact bodies within it to first post-Newtonian order. We find that our theory-independent parameters capture the behaviour of this full theory, and that it reduces to known scalar-tensor and vector-tensor limits in a well-behaved way. 

The ultimate use for our theory-independent parameters is for them to be constrained by gravitational wave data. However, the phase and amplitude of gravitational waves (at least for the inspiral) depend not only on the orbital dynamics of a binary but also on its radiative dynamics. When we constructed our phase and amplitude expressions in Reference \cite{pitt2025constraining}, we were reliant on a phenomenological parametrisation of the gravitational wave flux. In this regard, it would be of substantial benefit to have a fully-fledged parametrised post-Minkowskian formalism to describe the radiative sector together with conservative dynamics. This would afford two advantages: firstly, such parameters would, in principle, be better physically motivated. Secondly, it ought to be possible to map the new set of parameters to the existing PPN parameters, leading to a reduction in the total number of parameters in the phase and amplitude, potentially resulting in better constraints on deviations from GR. We leave the development of such a formalism to future work.

\vspace{0.5cm}
\noindent
{\bf Acknowledgements:}
Calculations were performed using the Mathematica package xAct. We acknowledge support from STFC under project reference 2897578.

\section*{Appendix A: Modified geodesic equation}
\label{sc:AppA}

This calculation follows the scalar sensitivity example from Reference \cite{pitt2025constraining} quite closely, the principal difference being the choice of quantity to which the mass of a particle is sensitive, i.e. that it is now $\gamma = -K^\mu u_\mu$. In this case, the mass's dependence on its own four-velocity leads to a different type of geodesic equation. Equation (\ref{eq:geo1}) is obtained by varying the following action:
\begin{align}
S_M 
&= - \sum_A \int 
    m_A(\gamma) \, 
    \big( -g_{\mu\nu} u^{\mu} u^{\nu} \big)^{1/2} 
    \, d\lambda \, ,
\end{align}
where $\lambda$ is a parameter along the curve, such that we can understand $\left( -g_{\mu\nu} u^{\mu} u^{\nu} \right)^{1/2} = {d\tau_A}/{d\lambda}$ where $\tau_A$ is the proper time along body A's world-line. We can expand around the background world-line to first order in $\delta x^\alpha$ to obtain
\begin{align}
\frac{d\tau_A}{d\lambda} 
&= \left( \frac{d\tau_A}{d\lambda} \right)_0 
- \frac{1}{2} g_{\mu\nu,\alpha} u_A^{\mu} u_A^{\nu} 
\left( \frac{d\tau_A}{d\lambda} \right)_0 \delta x_A^{\alpha} 
- g_{\alpha\nu} u_A^{\nu} \frac{d}{d\lambda} \delta x_A^{\alpha}, \\[1.5ex]
m_A(\gamma) 
&= m_A + m'_A \bigg( 
\frac{\partial \gamma}{\partial u_A^{\mu}} \delta u_A^{\mu} 
+ \frac{\partial \gamma}{\partial g_{\mu\nu}} \delta g_{\mu\nu} 
+ \frac{\partial \gamma}{\partial K^{\mu}} \delta K^{\mu} 
\bigg),
\end{align}
where $m_A = m_A(\gamma_0)$ and $m'_A = {\partial m_A}/{\partial \gamma_0}$, and where the subscript $0$ denotes evaluation along the unperturbed paths. We note also that
\begin{align}
\delta u_A^{\mu} 
&= \left( \frac{d\lambda}{d\tau_A} \right)_0 
\left( \delta^{\mu}_{\alpha} + u_A^{\mu} u_{A\alpha} \right)
\frac{d}{d\lambda} \delta x_A^{\alpha}  + \frac{1}{2} g_{\beta\gamma,\alpha} 
u_A^{\beta} u_A^{\gamma} u_A^{\mu} \delta x_A^{\alpha}, \\[1.2ex]
\delta g_{\mu\nu} 
&= g_{\mu\nu,\alpha} \delta x_A^{\alpha}, \qquad {\rm and} \qquad \delta K^{\mu} = K^{\mu}{}_{,\alpha} \, \delta x_A^{\alpha}.
\end{align}
We may use these results to expand $S_M$ to first order in deviations from a body's world-line and then, by demanding that $\delta S_M=0$, obtain the following equation of motion: 
\begin{align}
u_A^{\nu} \nabla_{\nu} \bigg[
m_A u_{A\alpha} 
- m'_A \frac{\partial \gamma}{\partial u_A^{\mu}}
\left( \delta^{\mu}_{\alpha} + u_A^{\mu} u_{A\alpha} \right)
\bigg]
&= - m'_A \bigg[
\frac{\partial \gamma}{\partial K^{\mu}} \, K^{\mu}{}_{,\alpha}
+ \frac{\partial \gamma}{\partial g_{\mu\nu}} \, g_{\mu\nu,\alpha}
- \frac{\partial \gamma}{\partial u_A^{\mu}} \, \Gamma^{\mu}_{\beta\alpha} u^{\beta}
\bigg].
\end{align}
This result then simplifies to give Equation (\ref{eq:geo1}).

\section*{Appendix B: The Modified EIH Formalism}

Einstein, Infeld and Hoffman originally introduced their Lagrangian as a
post-Newtonian description of interacting point masses in a gravitational field \cite{einstein1938gravitational}. In the modified EIH approach, the coefficients appearing in this Lagrangian are promoted to a set of parameters \cite{will2018testing}: 
\begin{align}
\left\{
\mathcal{A}_{a},
\mathcal{G}_{ab},
\mathcal{B}_{ab},
\mathcal{C}_{ab},
\mathcal{E}_{ab},
\mathcal{D}_{abc}
\right\},
\end{align}
which encode both the choice of gravitational theory and the internal properties of the bodies. The index symmetries of these coefficients are fixed by the
body-exchange symmetries of the corresponding terms in the Lagrangian. If, in
addition, one identifies the passive and active gravitational masses of each body,
these symmetries imply
\begin{equation}
    \mathcal{G}_{ab} = \mathcal{G}_{(ab)}, \qquad
    \mathcal{C}_{ab} = \mathcal{C}_{(ab)}, \qquad
    \mathcal{E}_{ab} = \mathcal{E}_{(ab)}, \qquad
    \mathcal{D}_{abc} = \mathcal{D}_{a(bc)}.
\end{equation}
One may note that $\mathcal{B}_{ab}$ is not in general symmetric due to it encoding preferred frame effects. In terms of these parameters, the modified-EIH Lagrangian has the form \cite{will2018testing}
\begin{equation}
\label{eq:EIHL}
\begin{aligned}
    L_{\text{EIH}} = & - \sum_{a} m_{a} \left[ 1 - \frac{1}{2} v_{a}^{2} - \frac{1}{8} (1 + \mathcal{A}_{a}) v_{a}^{4} \right] \\
    & + \frac{1}{2} \sum_{a} \sum_{b \neq a} \frac{m_{a} m_{b}}{r_{ab}} \left[ \mathcal{G}_{ab} + 3 \mathcal{B}_{ab} v_{a}^{2} - \frac{1}{2}(\mathcal{G}_{ab} + 6 \mathcal{B}_{(ab)} + \mathcal{C}_{ab}) \mathbf{v}_{a} \cdot \mathbf{v}_{b} \right. \\
    & \left. - \frac{1}{2} (\mathcal{G}_{ab} + \mathcal{E}_{ab}) (\mathbf{v}_{a} \cdot \mathbf{n}_{ab}) (\mathbf{v}_{b} \cdot \mathbf{n}_{ab}) \right]  - \frac{1}{2} \sum_{a} \sum_{b \neq a} \sum_{c \neq a} \mathcal{D}_{abc} \frac{m_{a} m_{b} m_{c}}{r_{ab} r_{ac}}.
\end{aligned}
\end{equation}
To obtain the equation of motion for a system of sensitive bodies in the non-preferred frame one can deploy a post-Galilean transformation to the EIH Lagrangian through a boost velocity $\mathbf{w}$. This gives rise to the following co-ordinate transformations:
\begin{align}
\mathbf{x} &= \boldsymbol{\xi}
+ \left(1 + \frac{1}{2} \mathrm{w}^{2}\right)\mathbf{w}\,\tau
+ \frac{1}{2}(\boldsymbol{\xi}\cdot\mathbf{w})\,\mathbf{w}
+ \boldsymbol{\xi}\times O(\epsilon^{2}),
\\[6pt]
t &= \tau\left(1 + \frac{1}{2} \mathrm{w}^{2} + \frac{3}{8} \mathrm{w}^{4}\right)
+ \left(1 + \frac{1}{2} \mathrm{w}^{2}\right)\boldsymbol{\xi}\cdot\mathbf{w}
+ \tau\times O(\epsilon^{3}) \, .
\end{align}
The equation of motion for such a boosted Lagrangian is given as \cite{will2018testing}
\begin{align}
\boldsymbol{a}_{1}
={}& -\frac{m_{2}\boldsymbol{n}}{r^{2}}
\Bigg\{
\mathcal{G}_{12}
-\left(3\mathcal{G}_{12}\mathcal{B}_{12}
+\mathcal{D}_{122}\right)\frac{m_{2}}{r}
\nonumber\\
&\quad
-\frac{1}{2}
\left[
2\mathcal{G}_{12}^{2}
+6\mathcal{G}_{12}\mathcal{B}_{(12)}
+2\mathcal{D}_{211}
+\mathcal{G}_{12}\left(\mathcal{C}_{12}
+\mathcal{E}_{12}\right)
\right]\frac{m_{1}}{r}
\nonumber\\
&\quad
+\frac{1}{2}
\left[
3\mathcal{B}_{12}
-\mathcal{G}_{12}\left(1+\mathcal{A}_{1}\right)
\right]v_{1}^{2}
+\frac{1}{2}
\left(
3\mathcal{B}_{21}
+\mathcal{G}_{12}
+\mathcal{E}_{12}
\right)v_{2}^{2}
\nonumber\\
&\quad
-\frac{1}{2}
\left(
6\mathcal{B}_{(12)}
+2\mathcal{G}_{12}
+\mathcal{C}_{12}
+\mathcal{E}_{12}
\right)\boldsymbol{v}_{1}\cdot\boldsymbol{v}_{2}
-\frac{3}{2}
\left(
\mathcal{G}_{12}
+\mathcal{E}_{12}
\right)
\left(\boldsymbol{n}\cdot\boldsymbol{v}_{2}\right)^{2}
\nonumber\\
&\quad
+\frac{1}{2}
\left(
\mathcal{C}_{12}
+\mathcal{G}_{12}\mathcal{A}_{1}
\right)w^{2}
+\frac{1}{2}
\left(
\mathcal{C}_{12}
-6\mathcal{B}_{[12]}
+\mathcal{E}_{12}
+2\mathcal{G}_{12}\mathcal{A}_{1}
\right)\boldsymbol{v}_{1}\cdot\boldsymbol{w}
\nonumber\\
&\quad
+\frac{1}{2}
\left(
\mathcal{C}_{12}
+6\mathcal{B}_{[12]}
-\mathcal{E}_{12}
\right)\boldsymbol{v}_{2}\cdot\boldsymbol{w}
+\frac{3}{2}\mathcal{E}_{12}
\left[
\left(\boldsymbol{w}\cdot\boldsymbol{n}\right)^{2}
+2\left(\boldsymbol{w}\cdot\boldsymbol{n}\right)
\left(\boldsymbol{v}_{2}\cdot\boldsymbol{n}\right)
\right]
\Bigg\}
\nonumber\\
&\quad
+\frac{m_{2}\boldsymbol{v}_{1}}{r^{2}}
\,\boldsymbol{n}\cdot
\left\{
\left[
3\mathcal{B}_{12}
+\mathcal{G}_{12}\left(1+\mathcal{A}_{1}\right)
\right]\boldsymbol{v}_{1}
-3\mathcal{B}_{12}\boldsymbol{v}_{2}
+\mathcal{G}_{12}\mathcal{A}_{1}\boldsymbol{w}
\right\}
\nonumber\\
&\quad
-\frac{1}{2}\frac{m_{2}\boldsymbol{v}_{2}}{r^{2}}
\,\boldsymbol{n}\cdot
\left\{
\left(
6\mathcal{B}_{(12)}
+2\mathcal{G}_{12}
+\mathcal{C}_{12}
+\mathcal{E}_{12}
\right)\boldsymbol{v}_{1}
\right.
\nonumber\\
&
\left.
-\left(
6\mathcal{B}_{(12)}
+\mathcal{C}_{12}
-\mathcal{E}_{12}
\right)\boldsymbol{v}_{2}
+2\mathcal{E}_{12}\boldsymbol{w}
\right\}
\nonumber\\
&\quad
-\frac{1}{2}\frac{m_{2}\boldsymbol{w}}{r^{2}}
\,\boldsymbol{n}\cdot
\left\{
\left(
\mathcal{C}_{12}
-6\mathcal{B}_{[12]}
+\mathcal{E}_{12}
-2\mathcal{G}_{12}\mathcal{A}_{1}
\right)\boldsymbol{v}_{1}
\right.
\nonumber\\
&
\left.
-\left(
\mathcal{C}_{12}
-6\mathcal{B}_{[12]}
-\mathcal{E}_{12}
\right)\boldsymbol{v}_{2}
-2\left(
\mathcal{G}_{12}\mathcal{A}_{1}
-\mathcal{E}_{12}
\right)\boldsymbol{w}
\right\},
\nonumber\\[1ex]
\boldsymbol{a}_{2}
={}& \left\{1 \rightleftharpoons 2;\, \boldsymbol{n}\to -\boldsymbol{n}\right\}.
\end{align}
One can insert the definitions of our new parameters, from Equations (\ref{eq:newparams}), to find the equation of motion in a non-preferred frame. It can also be seen that theories without preferred-frame effects satisfy the following \cite{will2018testing}:
\begin{equation}
    \mathcal{A}_a = \mathcal{B}_{[ab]} = \mathcal{C}_{ab} =\mathcal{E}_{ab} = 0.
    \label{eq:LIrequirements}
\end{equation}
For further details the reader is referred to Reference \cite{will2018theory}.

\section*{Appendix C: Gravitational Potentials}

One can define Newtonian and post-Newtonian potentials using the following functionals:
\begin{equation}
\begin{aligned}
P(f) &\equiv \frac{1}{4\pi}
\int_{\mathcal{M}} 
\frac{f(t,\mathbf{x}')}{|\mathbf{x}-\mathbf{x}'|}
\, d^3x', \\[6pt]
\Sigma(f)
&\equiv \int_{\mathcal{M}}
\frac{\rho^{*}(t,\mathbf{x}')\, f(t,\mathbf{x}')}{|\mathbf{x}-\mathbf{x}'|}
\, d^{3}x'
= P\!\left(4\pi \rho^{*} f\right), \\[6pt]
\Sigma_{s}(f)
&\equiv \int_{\mathcal{M}}
\frac{(1-2s_\phi(\mathbf{x}'))\,\rho^{*}(t,\mathbf{x}')\, f(t,\mathbf{x}')}
{|\mathbf{x}-\mathbf{x}'|}
\, d^{3}x'
= P\!\left(4\pi (1-2s_\phi)\rho^{*} f\right)\\[6pt]
\Sigma_\gamma (f)
&\equiv \int_{\mathcal{M}}
\frac{(1-s_\gamma(\mathbf{x}'))\,\rho^{*}(t,\mathbf{x}')\, f(t,\mathbf{x}')}
{|\mathbf{x}-\mathbf{x}'|}
\, d^{3}x'
= P\!\left(4\pi (1-s_\gamma)\rho^{*} f\right)\\[6pt]
X(f) 
&\equiv \int_{\mathcal{M}}
\rho^{*}(t,\mathbf{x}')\, f(t,\mathbf{x}') 
\, |\mathbf{x}-\mathbf{x}'| \, d^{3}x' \,.
\end{aligned}
\end{equation}
The Newtonian and the post-Newtonian potentials are then given by
\begin{equation}
\begin{alignedat}{3}
U &\equiv \Sigma(1)
&\qquad U_s &\equiv \Sigma_s(1)
&\qquad V^{j} &\equiv \Sigma_\gamma(v^j)
\\[6pt]
\Phi_{1} &\equiv \Sigma_\gamma(v^2)
&\qquad \Phi^{s}_{1} &\equiv \Sigma(v^{2}(1-s_\gamma-2s_\phi+2s_{\gamma\phi}-2s_\gamma s_\phi))
&\qquad \Phi_{2} &\equiv \Sigma(U)
\\[6pt]
\Phi^{s}_{2} &\equiv \Sigma_{s}(U)
&\qquad \Phi_{2s} &\equiv \Sigma(U_{s})
&\qquad \Phi^{s}_{2s} &\equiv \Sigma_{s}(U_{s})
\\[6pt]
X &\equiv X(1)
&\qquad X_{s} &\equiv X(1-2s_\phi)
&\qquad X_{\ae}^j &\equiv X(s_\gamma v^j)
\\[6pt]
V^j_{\ae} &\equiv \Sigma(s_\gamma v^j)\,,
\end{alignedat}
\end{equation}
which can further be used to define 
\begin{equation}
\begin{aligned}
U_{s\sigma}
&= U_{s} + \epsilon \left\{
-\frac{1}{2}\Phi^{s}_{1}
- 3 G(1-\zeta)\Phi^{s}_{2}
+ 3 G\zeta \Phi^{s}_{2s}
- 4\alpha G\zeta \Sigma(a_{\phi}U_{s})
\right\},\\[6pt]
U_{\sigma}
&= U + \epsilon \left\{
\frac{3}{2}\Phi_{1}
- G(1-\zeta)\Phi_{2}
+ 6\alpha G\zeta \Phi_{2s}
- \alpha G \zeta \Phi^{s}_{2s}\right\},\\[6pt]
V^{i}_{\sigma} &= V^{i}, \\[6pt]
V^{i}_{\ae\,\sigma} &= -\,V^{i}_{\ae}, \\[6pt]
X_{\sigma} &= X , \\[6pt]
X^{i}_{\ae\,\sigma} &= -\,X^{i}_{\ae}\,.
\end{aligned}
\end{equation}
These potentials are used in Section \ref{sc:FullTheory} of this paper.

\section*{References}
\bibliographystyle{ieeetr}


\end{document}